\documentclass[oldversion]{aa}
\usepackage{epsfig}
\usepackage{natbib}
\usepackage{lscape}
\usepackage{amssymb}
\usepackage{epstopdf}
\usepackage[caption=false]{subfig}
\usepackage[caption=false]{subfig}
\usepackage{float}

\usepackage{longtable}
\usepackage{color}
\usepackage{fancyheadings}
\bibpunct{(}{)}{;}{a}{}{,}

\begin{document}

\title{Activity induced variation in spin-orbit angles as derived from Rossiter-McLaughlin measurements\thanks{RV measurements obtained from the HARPS pipeline are available in electronic form at the CDS via anonymous ftp to cdsarc.u-strasbg.fr (130.79.128.5) or via http://cdsarc.u-strasbg.fr/viz-bin/qcat?J/A+A/606/A107}}

\author{M. Oshagh\inst{1,2}, A. H. M. J. Triaud\inst{3}, A. Burdanov\inst{4}, P. Figueira\inst{2,5}, A. Reiners\inst{1}, N. C. Santos\inst{2,6}, J. Faria\inst{2}, G. Boue\inst{7}, R. F. D\'iaz\inst{8,9}, S. Dreizler\inst{1}, S. Boldt\inst{1}, L. Delrez \inst{10}, E. Ducrot\inst{4}, M. Gillon\inst{4}, A. Guzman Mesa\inst{1}, E. Jehin\inst{4}, S. Khalafinejad\inst{11,12}, S. Kohl\inst{11}, L. Serrano\inst{2}, S. Udry\inst{13} }

\institute{Institut f\"ur Astrophysik, Georg-August Universit\"at G\"ottingen,
Friedrich-Hund-Platz 1, 37077 G\"ottingen, Germany
\and
Instituto de Astrof\' isica e Ci\^encias do Espa\c{c}o, Universidade do Porto, CAUP, Rua das Estrelas, PT4150-762 Porto, Portugal 
\and
University of Birmingham, Edgbaston, Birmingham, B15 2TT, UK
\and
Space sciences, Technologies and Astrophysics Research (STAR) Institute, Universit\'e de Li\`ege,  All\'ee du 6 Ao\^ut 17,  Bat.  B5C, 4000 Li\`ege, Belgium
\and
European Southern Observatory, Alonso de Cordova 3107, Vitacura Casilla 19001, Santiago 19, Chile
\and
Departamento de F{\'i}sica e Astronomia, Faculdade de Ci{\^e}ncias, Universidade do
Porto,Rua do Campo Alegre, 4169-007 Porto, Portugal
\and
IMCCE, Observatoire de Paris, UPMC Univ. Paris 6, PSL Research University, Paris, France
\and
Universidad de Buenos Aires, Facultad de Ciencias Exactas y Naturales, Buenos Aires, 1428, Argentina 
\and
CONICET - Universidad de Buenos Aires, Instituto de Astronomía y F\'isica del Espacio (IAFE), Buenos Aires, 1428, Argentina
\and
Astrophysics Group, Cavendish Laboratory, J. J. Thomson Avenue, Cambridge, CB3 0HE, UK
\and
Hamburg Observatory, Hamburg University, Gojenbergsweg 112, 21029, Hamburg, Germany 
\and
Max Planck Institute for Astronomy, K{\"o}nigstuhl 17, 69117 Heidelberg, Germany
\and
Observatoire de Gen\`{e}ve, Universit\'{e} de Gen\`{e}ve, 51 chemin des Maillettes, 1290 Versoix, Switzerland
}

\date{Received XXX; accepted XXX}

\abstract {One of the most powerful methods used to estimate sky-projected spin-orbit angles of exoplanetary systems is through a spectroscopic transit observation known as the Rossiter–McLaughlin (RM) effect. So far mostly single RM observations have been used to estimate the spin-orbit angle, and thus there have been no studies regarding the variation of estimated spin-orbit angle from transit to transit. Stellar activity can alter the shape of photometric transit light curves and in a similar way they can deform the RM signal. In this paper we discuss several RM observations, obtained using the HARPS spectrograph, of known transiting planets that all transit extremely active stars, and by analyzing them individually we assess the variation in the estimated spin-orbit angle. Our results reveal that the estimated spin-orbit angle can vary significantly  (up to $\sim 42^\circ$) from transit to transit, due to variation in the configuration of stellar active regions  over different nights. This finding is almost two times larger than  the expected variation predicted from simulations. We could not identify any meaningful correlation between the variation of estimated spin-orbit angles and the stellar magnetic activity indicators. We also investigated two possible approaches to mitigate the stellar activity influence on RM observations. The first strategy was based on obtaining several RM observations and folding them to reduce the stellar activity noise. Our results demonstrated that this is a feasible and robust way to overcome this issue. The second approach is based on acquiring simultaneous high-precision short-cadence photometric transit light curves using TRAPPIST/SPECULOOS telescopes, which provide more information about the stellar active region's properties and allow a better RM modeling.
}

\keywords{methods: observational, numerical- planetary system- techniques: photometry, spectroscopy}

\authorrunning{M. Oshagh et al.}
\titlerunning{Activity induced variation in measured spin-orbit angle}
\maketitle
\section{Introduction}

Understanding the orbital architecture of exoplanetary systems is one of the main objectives of
exoplanetary surveys, in order to provide new strong constraints on the physical mechanisms behind
planetary formation and evolution. The sky-projected spin-orbit angle (hereafter spin-orbit angle $\lambda$)\footnote{The angle between the stellar spin axis and normal to the planetary orbital plane.} is one of the most important parameters of the orbital architecture of planetary
systems. According to the core-accretion paradigm of planet formation, planets should form on aligned orbits $\lambda \sim 0$  \cite[\citealp{Pollack-96}, and see a comprehensive review in][and references therein ]{Dawson-18}, as observed in the case of the solar system. Planetary evolution mechanisms such as disk migration predict that planets should maintain their primordial spin-orbit angle, but  the dynamical interactions with a third body and tidal migration could alter the spin-orbit angle \cite[see a comprehensive review in][and references therein]{Baruteau-14}. Moreover, several studies have shown that the misalignment of a planet could be primordial considering the disks being naturally and initially misaligned \citep[e.g.,][]{Batygin-12, Crida-14}. In this context, the measurement of spin-orbit angles under a wide range of host and planet conditions (including multiple transiting planets) is extremely important in order to obtain feedback on modeling.

As a star rotates, the part of its surface that rotates toward the
observer is blueshifted and the part that rotates away is redshifted. During the transit of a planet, the corresponding
rotational velocity of the portion of the stellar disk that is
blocked by the planet is removed from the integration of the
velocity over the entire star, creating a radial velocity (RV)
signal  known as the Rossiter--McLaughlin (RM) effect \citep{Holt-1893, Rossiter-24, McLaughlin-24}. The RM observation is the
 most powerful and efficient technique for  estimating the spin-orbit angle $\lambda$ of exoplanetary systems (or  eclipsing binary systems) \cite[e.g., \citealp{Winn-05, Hebrard-08, Triaud-10, Albrecht-12}, and a comprehensive review in][and references therein]{Triaud-18}.  The number of spin-orbit angle $\lambda$ measurements has dramatically increased over the past decade (200 systems to date)\footnote{http://www.astro.keele.ac.uk/jkt/tepcat/obliquity.html}, and have revealed a large number of unexpected systems, which include planets on
highly misaligned \citep[e.g.,][]{Triaud-10, Albrecht-12, Santerne-14}, polar \citep[e.g.,][]{Addison-13}, and even retrograde orbits \citep[e.g.,][]{Bayliss-10, Hebrard-11}.

\begin{table*}
\caption{Planetary and stellar parameters of our targets}              
\centering                                      
\begin{tabular}{c c c c c c c c c }          
\hline\hline                        
Parameter & Symbol & Unit & WASP-6b & WASP-19b & WASP-41b & WASP-52b & CoRoT-2b & Qatar-2b \\
\hline 
Stellar radius & $R_{\star}$ & $R_{\odot}$ & 0.87 & 1.004 & 1.01&       0.79&         0.902 & 0.713\\

Planet-to-star radius ratio & $R_{p}/R_{\star}$ & - & 0.1463 & 0.1488 & 0.13674 & 0.16462& 0.1667 & 0.16208 \\
Scaled semimajor axis & $a/R_{\star}$ & - & 10.4085 & 3.552 & 9.96 &7.38 & 6.70 &  5.98\\
Orbital inclination & $i$ & $^\circ$ & 88.47 &78.94 & 88.7 &85.35 & 87.84 &  86.12\\
Orbital period & $P$ & days & 3.3610060 & 0.78884& 3.0524040 & 1.7497798 & 1.7429935 & 1.33711647 \\
Linear limb darkening & $u_{1}$ &- & 0.386& 0.427 & 0.3 &0.62& 0.346 & 0.6231\\
Quadratic limb darkening & $u_{2}$ &- & 0.214& 0.222 &0.25 & 0.06  &0.220 & 0.062  \\

\hline                                             

\end{tabular}
\end{table*}

The contrast of active regions (i.e., stellar spots, plages, and faculae) present on the stellar surface during the transit of an exoplanet alter the high-precision photometric transit light curve (both outside and inside transit) and lead to an
inaccurate estimate of the planetary parameters \citep[e.g.,][]{Czesla-09, Sanchis-Ojeda-11b, Sanchis-Ojeda-13, Oshagh-13b, Barros-13, Oshagh-15a, Oshagh-15b, Ioannidis-16}. Since the
physics and geometry behind the transit light curve and the RM measurement are the same, the RM observations are expected to be affected by the presence of  stellar
active regions in a similar way. Furthermore, the inhibition of the convective blueshift inside the active regions can induce extra RV noise on the RM observations, which is absent in the photometric transit observation.

\citet{Boldt-18} demonstrate, using simulations, that unocculted stellar active regions during the planetary transit influence the RV slope of out-of-transit in RM observations (due to a combination of  contrast and stellar rotation) and leads to an inaccurate estimate of the planetary radius; however, their influence on the estimated $\lambda$ was negligible. On the other hand, \citet{Oshagh-16} demonstrated that the occulted stellar active regions can lead to quite significant inaccuracies in the estimation
of spin-orbit angle $\lambda$ (up to $\sim 30 ^\circ$ for Neptune-sized planets and up to $\sim 15 ^\circ$ for hot Jupiters), which is much larger than the typical error bars on the $\lambda$ measurement, especially in the
era of high-precision RV measurements like those provided
by HARPS \citep{Mayor-03}, HARPS-N \citep{Cosentino-12}, CARMENES \citep{Quirrenbach-14}, and that will be provided by ESPRESSO \citep{Pepe-14}.

\begin{table*}
\caption{Summary of RM observation nights, and the simultaneous photometry}              
\centering                                      
\begin{tabular}{c c c c c c c c}          
\hline\hline                        
Name & First RM & Second RM & Third RM & Extra RM & Simultaneous Phot\#1 &Simultaneous Phot\#2 \\    
\hline          
\\                         
WASP-6b & 2017-09-17 & NO & NO & 2008-10-18$^a$ & NO & NO \\
WASP-19b & 2017-05-05 & 2017-05-09 & NO & 2010-03-19$^a$ & 2017-05-05$^d$ & 2017-05-09$^d$\\
WASP-41b & 2017-04-15 & 2017-04-21 & NO & 2011-04-03$^a$ & NO & NO\\
WASP-52b & 2017-09-21 & NO & NO & 2011-08-21$^b$ & NO & NO\\
CoRoT-2b & 2017-07-12 & 2017-07-19& 2017-09-04 &2007-09-02$^a$ & 2017-07-12$^d$& 2017-07-19$^d$\\
Qatar-2b & 2017-07-10 & NO & NO  & 2014-04-27$^c$ & 2017-07-10$^{d,e}$  & NO\\
\hline                                             
\end{tabular}
\begin{flushleft} 
$^a$ HARPS\\
$^b$ SOPHIE\\
$^c$ HARPS-N\\
$^d$ TRAPPIST\\
$^e$ SPECULOOS\\
\end{flushleft}
\end{table*}

In almost all (all could be an overstated or exaggerated statement) studies of RM observations, only one RM observation during a single transit  was acquired, and single-epoch RM observations were used to estimate the spin-orbit angle $\lambda$ of the systems. The main reason for this is the high telescope pressure and oversubscription on high-precision spectrographs that makes requesting for several RM observations of same targets inaccessible. Therefore, the lack of multiple RM observations has not permitted us to explore possible contamination on the estimated $\lambda$ from the stellar activity.

In this paper we assess the impact of stellar activity on the spin-orbit angle estimation by observing several RM observations of known transiting planets which all transit extremely active stars. We organize our paper as follows. We describe our targets' selection and their observations in Sect. 2. In Sect. 3 we define our fitting procedure to estimate the spin-orbit angles, and also present its result. We investigate the possibility of mitigating the influence of  stellar activity  on RM observations by combining several RM observations in Sect. 4. In Sect. 5 we present the simultaneous photometric transit observation with our RM observation and we discuss the advantages of having simultaneous photometric transit observation in eliminating stellar activity noise in RM observations. In Sect. 6 we discuss possible contamination from other sources of stellar noise that affect our observations, and conclude  in Sect 7.

\begin{figure*}
\center
\vspace{-0.5cm}
  \subfloat{\includegraphics[width=0.5 \textwidth]{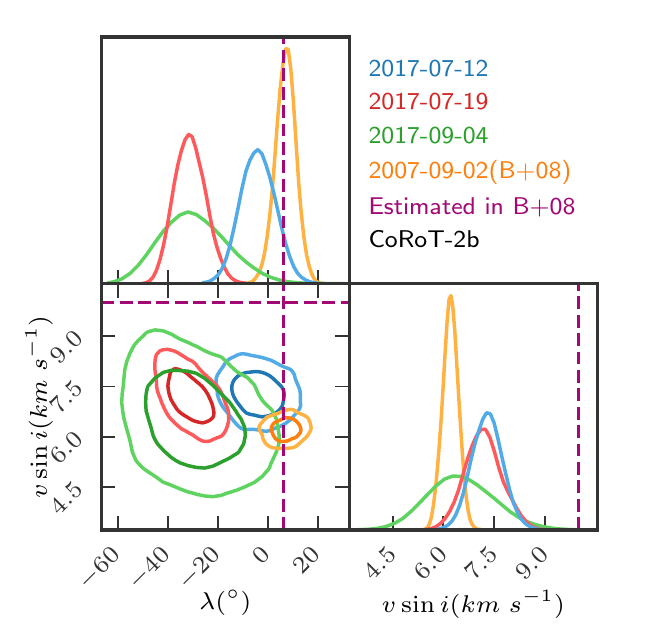}} \hspace{-0.2cm}\medskip
  \subfloat{\includegraphics[width=0.5 \textwidth]{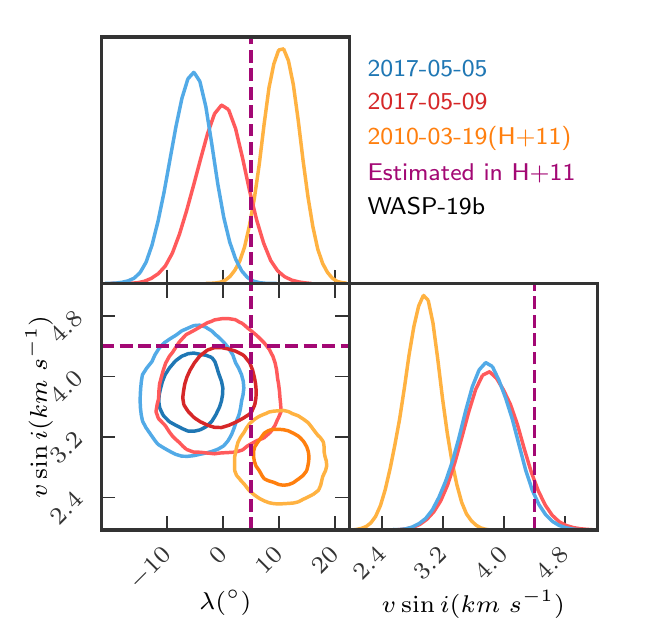}}
\medskip        \vspace{-1.7cm} 
\hspace{-1.3cm} \subfloat{\includegraphics[width=0.5 \textwidth]{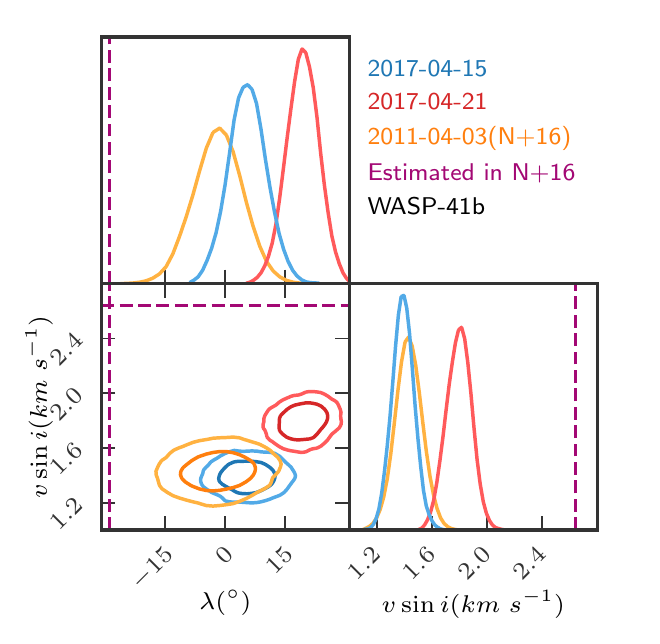}} \hspace{-0.2cm}\medskip
  \subfloat{\includegraphics[width=0.5 \textwidth]{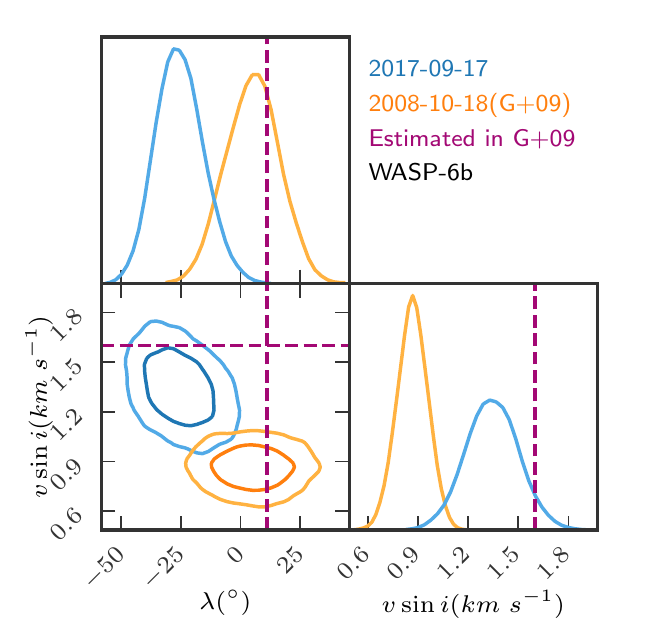}}
\medskip        \vspace{-1.7cm} \hspace{-1.3cm} 
 \subfloat{\includegraphics[width=0.5 \textwidth]{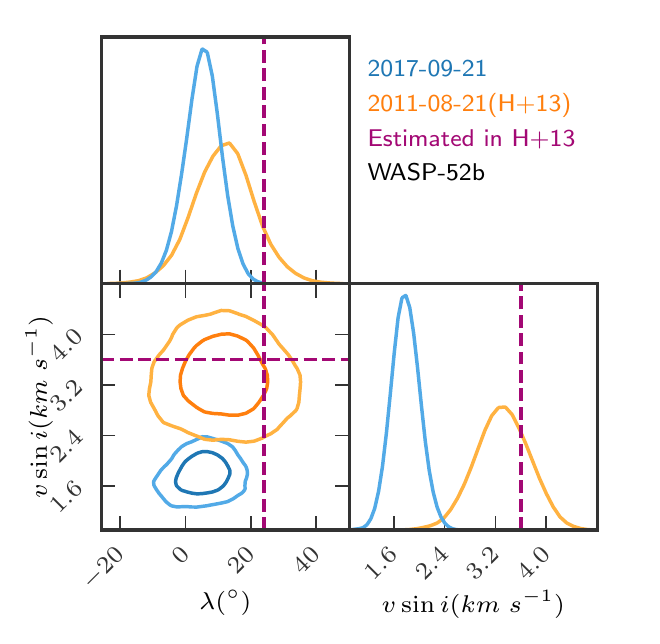}} \hspace{-0.2cm}\medskip
  \subfloat{\includegraphics[width=0.5 \textwidth]{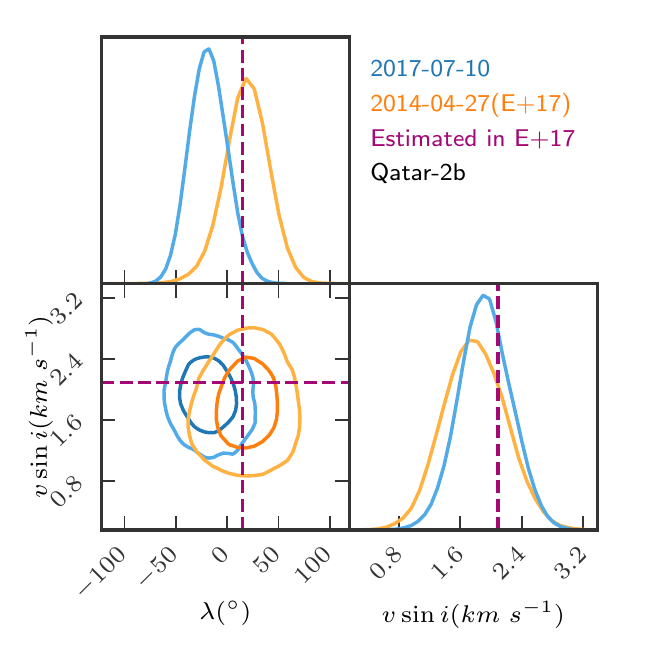}}

 \caption{Posterior probability distributions in  $v \sin i - \lambda $ parameter space of all our targets obtained from the fit to individual RM observations obtained at different transits and on different nights. Each panel shows different planetary system, and in each panel different colors correspond to the different nights on which the RM observation were performed. The purple dashed line displays the reported value in the literature. B+08: \citet{Bouchy-08}, H+11: \citet{Hellier-11}, N+16: \citet{Neveu-VanMalle-16}, G+09: \citet{Gillon-09}, H+13: \citet{Hebrard-13}, E+17: \citet{Esposito-17}.}

  \end{figure*}

\section{Observations and data reduction}
\subsection{Target selection}

Our stars were selected following four main criteria:

\begin{itemize}
\item{I) the star should have a known transiting planet;}
\item{II) the planet host star should be very active (e.g., the published photometric transit light curves  clearly show large stellar spots' occultation anomalies);}
\item{III) at least one RM measurement of the planet has been observed (preferably with the HARPS spectrograph as we  use HARPS observations in our study) to ensure the feasibility of detecting the RM signal;}
\item{IV) the transiting planet should have several observable transits from the southern hemisphere during the period from March 2017 to October 2017 (this criterion is to ensure observability).}
\end{itemize} 

Our criteria resulted in the selection of six transiting exoplanets, namely, WASP-6b, WASP-19b, WASP-41b, WASP-52b, CoRoT-2b, and Qatar-2b. Their published  transit light curves have indicated the presence of stellar spots on their surfaces with a filling factor of 6\%, 8\% , 3\%, 15\%, 16\%, and 4\%, respectively \citep{Tregloan-Reed-15, Sedaghati-15, Southworth-16, Mancini-17, Nutzman-11, Dai-17}. Most of them also have at least one observed RM with HARPS \citep{Gillon-09, Hellier-11, Neveu-VanMalle-16, Bouchy-08}, except for  Qatar-2, which has one observed RM with the HARPS-N spectrograph \citep{Esposito-17} and for WASP-52, which has one observed RM with the SOPHIE spectrograph \citep{Hebrard-13}.

We list all the planetary and stellar parameters of our targets collected from the literature and that are necessary for our analysis  in Table 1.

\subsection{HARPS observation}
Our program was allocated 60 hours on the HARPS spectrograph, mounted on the ESO 3.6 m telescope at La Silla observatory  \citep{Mayor-03}, to carry out high-precision RV measurements during three transits of each of our targets (under ESO programme ID: 099.C-0093, PI: M. Oshagh). The main aim of our program was to measure the spin-orbit angle $\lambda$ from individual RM observation of each target,  by having multiple $\lambda$ measurements for each target, and to quantify the changes in the measured $\lambda$ from transit to transit.

Thanks to the time-sharing scheme on HARPS with several other observing programs we were able to spread our 60 hours (equivalent to six nights) over a large fraction of the semester and obtain observations during several nights in which transits occur. Due to bad weather we lost almost 35\% of our allocated time, thus we could not obtain three RM observations for all of our targets. We summarize the number of RM observations collected
for each target and the nights  in Table 2. To overcome the shortage of several RM observations for some of our targets, we decided to collect the publicly available RM observations, preferably obtained with HARPS, of those targets (called extra RM). The extra RM observations, their observed dates, and  the spectrograph are also listed in Table 2.

The RM observation for each transit started at least one hour before the start of the transit  and lasted until one hour after the transit ended. The exposure times range from 900 to 1200 seconds,
ensuring a constant signal-to-noise ratio. All our targets throughout the observations remained above airmass 1.8 ($\textit{X} < 1.8$). The spectra were acquired with
simultaneous Fabry--Perot spectra on  fiber B for simultaneous wavelength reference, and were taken with the detector in fast-readout mode to minimize overheads and increase the total integration time during transit.

The collected HARPS spectra in our study were reduced using the HARPS Data Reduction Software (DRS - \citealt{Pepe-02, Lovis-07}). The spectra were cross-correlated with masks based on their stellar spectral type. As output the DRS provides the RVs and associated error.

\subsection{TRAPPIST/SPECULOOS observations}

We obtained additional simultaneous high-precision photometric transit observations with some of our RM observations using TRAPPIST-South\footnote{www.trappist.uliege.be} \citep{gillon2011, jehin2011}, and one of the SPECULOOS\footnote{www.speculoos.uliege.be} Southern Observatory telescopes \citep{gillon2018, burdanov2017}. TRAPPIST-South is a 60 cm (F/8) Ritchey--Chr\'etien telescope installed by the University of Li\`ege in 2010 at the ESO La Silla Observatory in the Atacama Desert in Chile. The SPECULOOS Southern Observatory is a facility composed of four robotic Ritchey--Chr\'etien (F/8) telescopes of 1 m diameter currently being installed at the ESO Paranal Observatory (PI of both telescopes: M. Gillon). The main objective of obtaining simultaneous high-precision and high-cadence photometric transit light curves was to assist us in identifying the presence of occultations of   active regions by the planet during the transits. We listed the nights on which simultaneous photometric transits were collected in Table 2.


The  observation for each transit started at least one hour before the
start of the transit  and lasted until one hour after the transit ended. The TRAPPSIT and SPECULOOS observations were acquired in V band and Sloan g', respectively. The exposure times range from 10 to 35 seconds.

Data reduction consisted of standard calibration steps (bias, dark, and flat-field corrections) and subsequent aperture photometry in IRAF/DAOPHOT \cite{Stetson1987}. Comparison stars and aperture size were selected manually to ensure the best photometric quality in terms of the flux standard deviation of check stars, i.e., non-variable stars similar in terms of magnitude and color to the target star.

\section{Variation of spin-orbit angle}
In this section we aim to determine the changes in measured spin-orbit angle $\lambda$ estimated from individual RM observations for a sample of exoplanets.

\begin{figure}[h!]
\includegraphics[width=0.45 \textwidth, height=8cm]{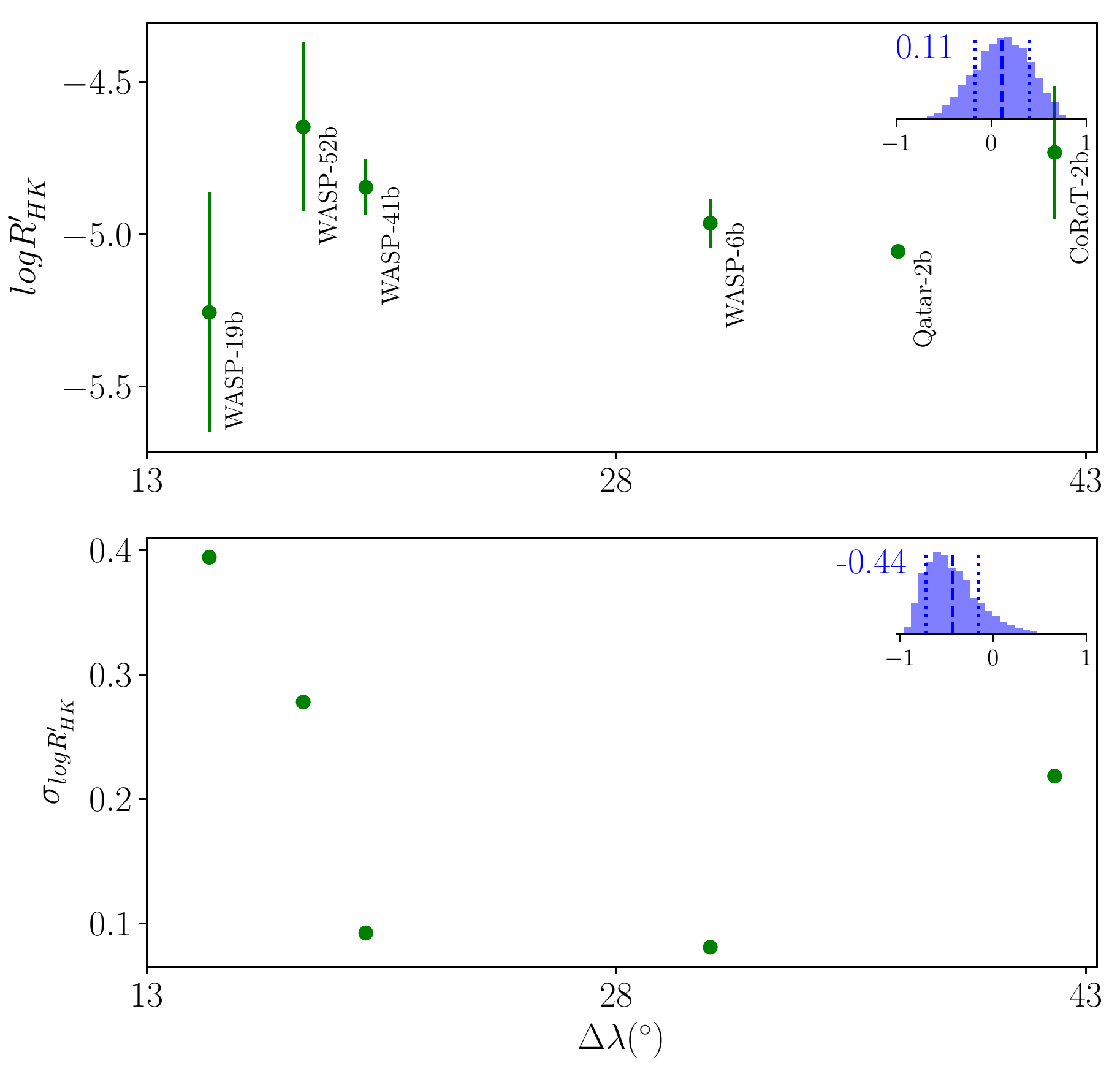}
\caption{Top: Correlations between $\Delta \lambda$ and $log R'_{HK}$. At the top right the calculated value of
$\rho$ and also its posterior distribution is presented. The dashed vertical lines indicate the 68\% highest posterior density
credible intervals \citep{Figueira-16}. Bottom: Same as top but for $\Delta \lambda$ and $\sigma_{log R'_{HK}}$. }
\end{figure}

\subsection{RM model}

To estimate the spin-orbit angle $\lambda$ from RM observations, we use the publicly available code \textit{ARoME}\footnote{www.astro.up.pt/resources/arome} \citep{Boue-12b}, which provides an analytical model to compute the RM signal, and is optimized for the spectrographs that utilize the CCF-based approach to estimate RVs (such as HARPS). \textit{ARoME} requires as input parameters the width of a non-rotating star which can be considered as the instrumental broadening profile ($\beta_{0}$), width of the best Gaussian fit to out-of-transit CCF ($\sigma_{0}$), stellar macroturbulence ($\zeta$), stellar radius ($R_{\star}$), projected stellar rotational velocity $v \sin i$, stellar quadratic limb darkening coefficients ($u_{1}$ and $u_{2}$), planetary semimajor axis ($a$), orbital period of planet ($P$), planet-to-star radius ratio ($R_{p}/R_{\star}$), orbital inclination angle ($i$), and the spin-orbit angle $\lambda$ in order to calculate the RM signal.

\citet{Brown-17} performed a comparison study between \textit{ARoME} and other RM modeling tools, and reached the conclusion that \textit{ARoME} consistently underestimates $v \sin i$ when compared to other models. Although $v \sin i$ differed, the estimated $\lambda$ were in strong agreement for all the models. Moreover, \textit{ARoME} is the only code which is publicly available, and  we are only interested in estimating $\lambda$ and in its variation, thus we decided to use the \textit{ARoME} tool for our analysis.

\begin{figure*}
\center
\vspace{-0.5cm}
  \subfloat{\includegraphics[width=0.45 \textwidth]{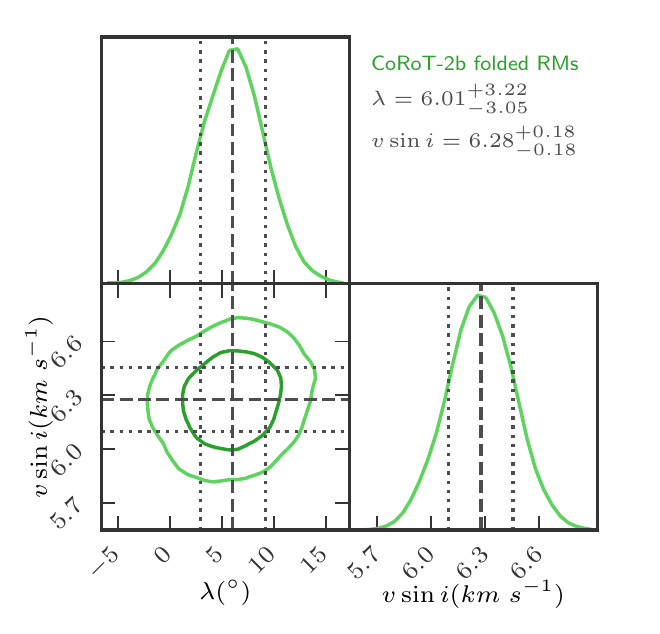}} \hspace{-0.2cm}\medskip
  \subfloat{\includegraphics[width=0.45 \textwidth]{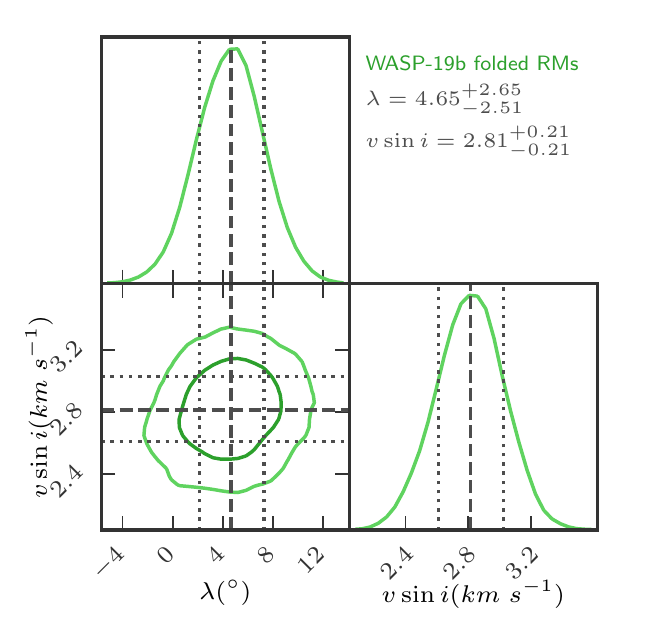}}
\medskip        \vspace{-1.1cm} 
\hspace{-1.3cm} \subfloat{\includegraphics[width=0.45 \textwidth]{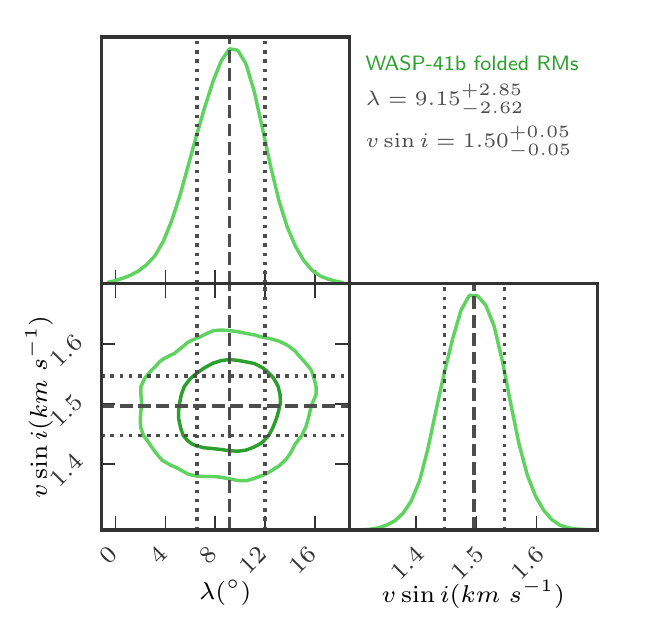}} \hspace{-0.2cm}\medskip
  \subfloat{\includegraphics[width=0.45 \textwidth]{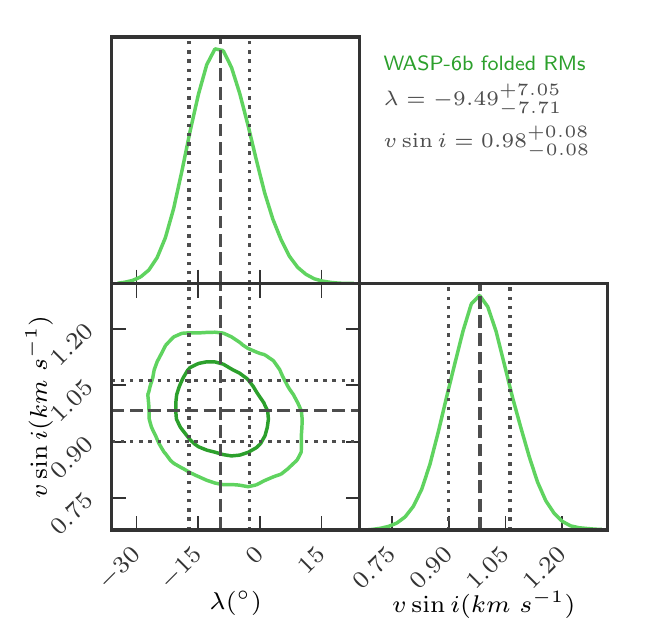}}
\medskip        \vspace{-1.1cm} \hspace{-1.3cm} 
 \subfloat{\includegraphics[width=0.45 \textwidth]{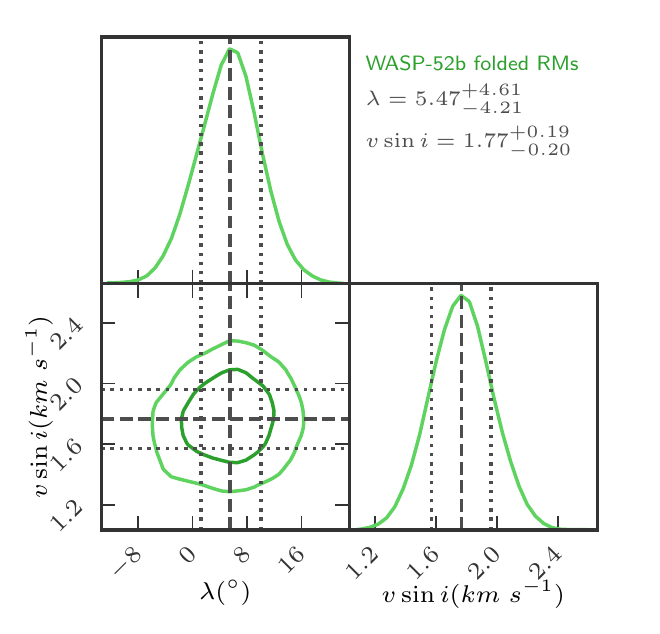}} \hspace{-0.2cm}\medskip
  \subfloat{\includegraphics[width=0.45 \textwidth]{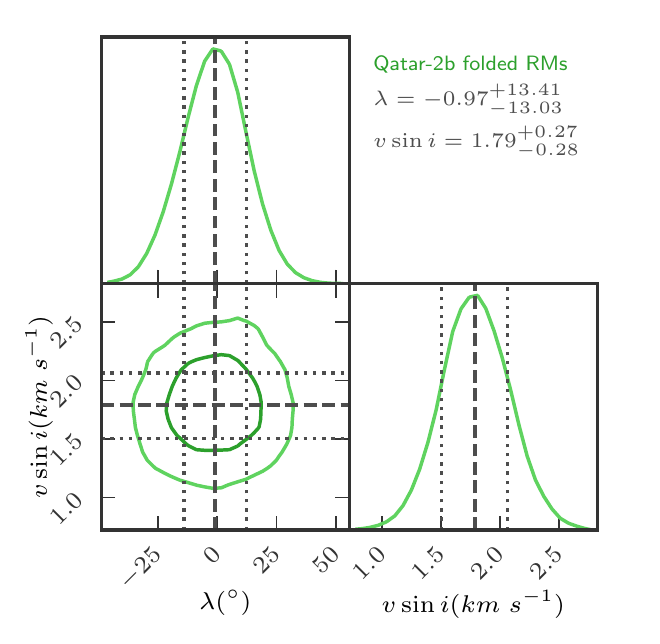}}
\caption{Posterior probability distributions in $v \sin i - \lambda$ parameter space of all our targets obtained from the fit to the folded RM observations. Each panel shows different planetary system. The black dashed line displays median values of the posterior distributions, and the $1 \sigma$ uncertainties taken to be the value enclosed in the 68.3 percent of the posterior distributions are shown with the black dotted line.}
\end{figure*}

\begin{figure*}[h!]
 \centering
\vspace{-0.cm}
  \subfloat{\includegraphics[width=0.5 \textwidth]{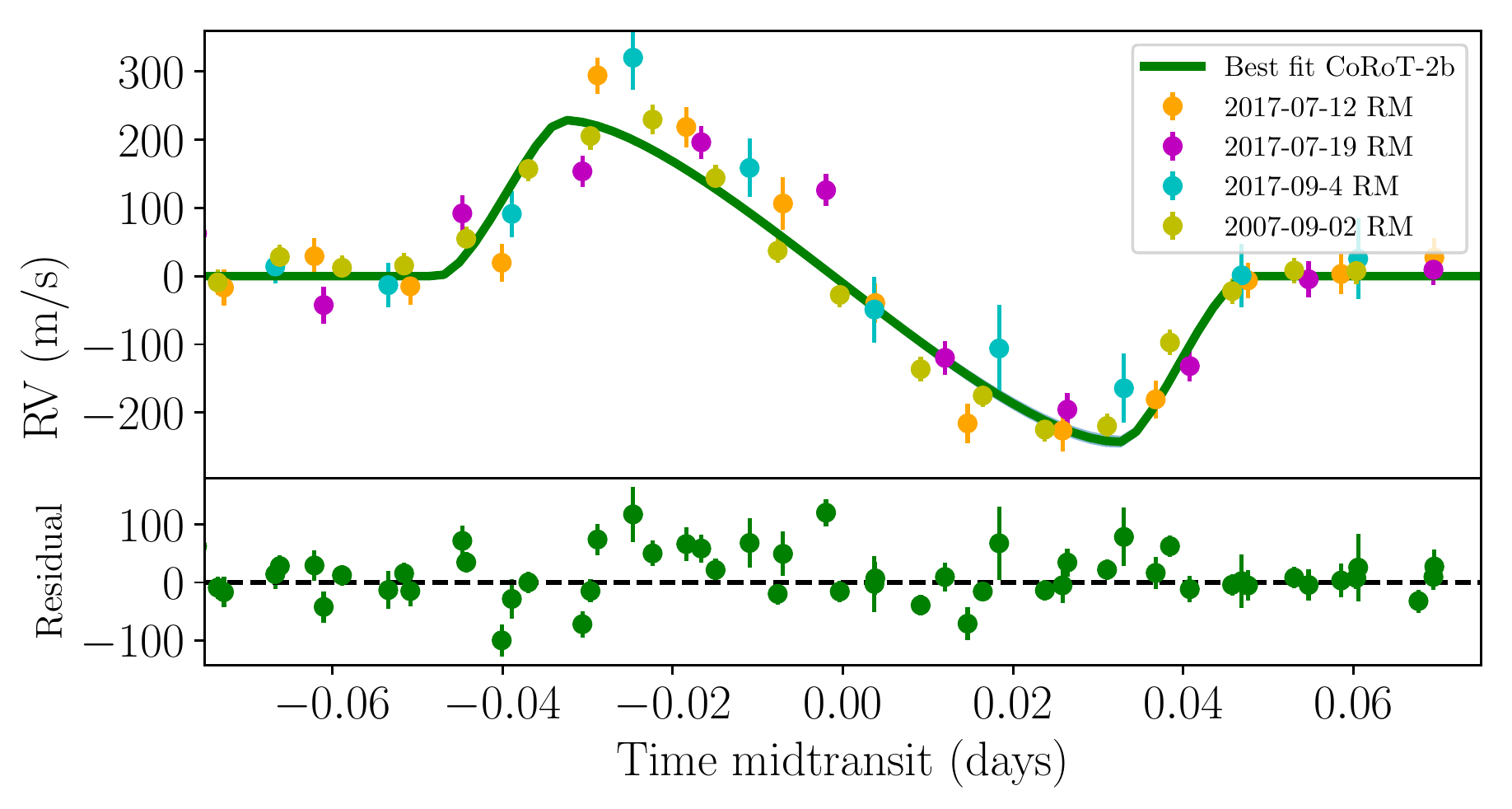}} \hspace{0.cm}\medskip
  \subfloat{\includegraphics[width=0.5 \textwidth]{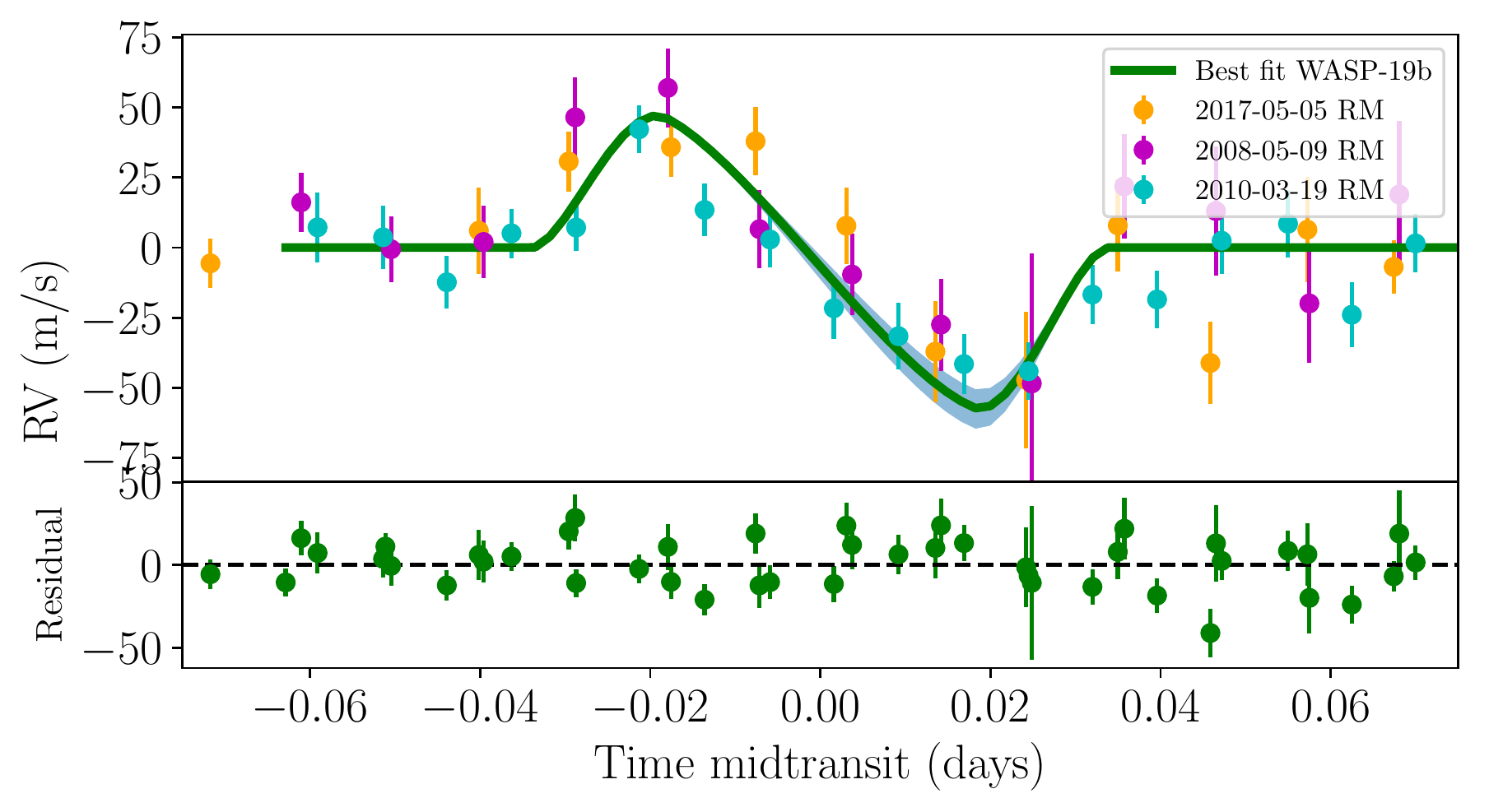}}
\medskip        \vspace{-1.cm} 
\hspace{0.cm} \subfloat{\includegraphics[width=0.5 \textwidth]{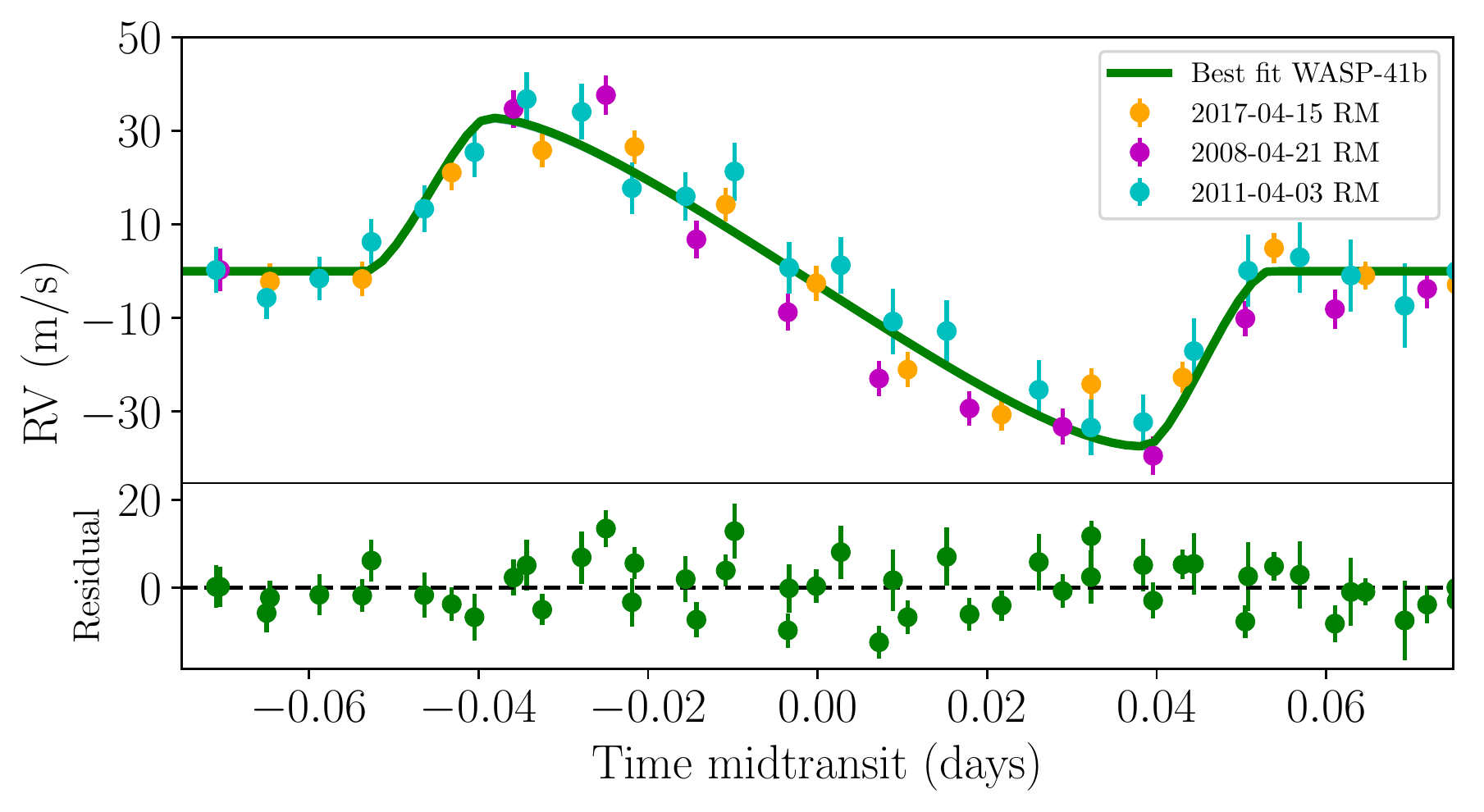}} \hspace{-0.cm}\medskip
  \subfloat{\includegraphics[width=0.5 \textwidth]{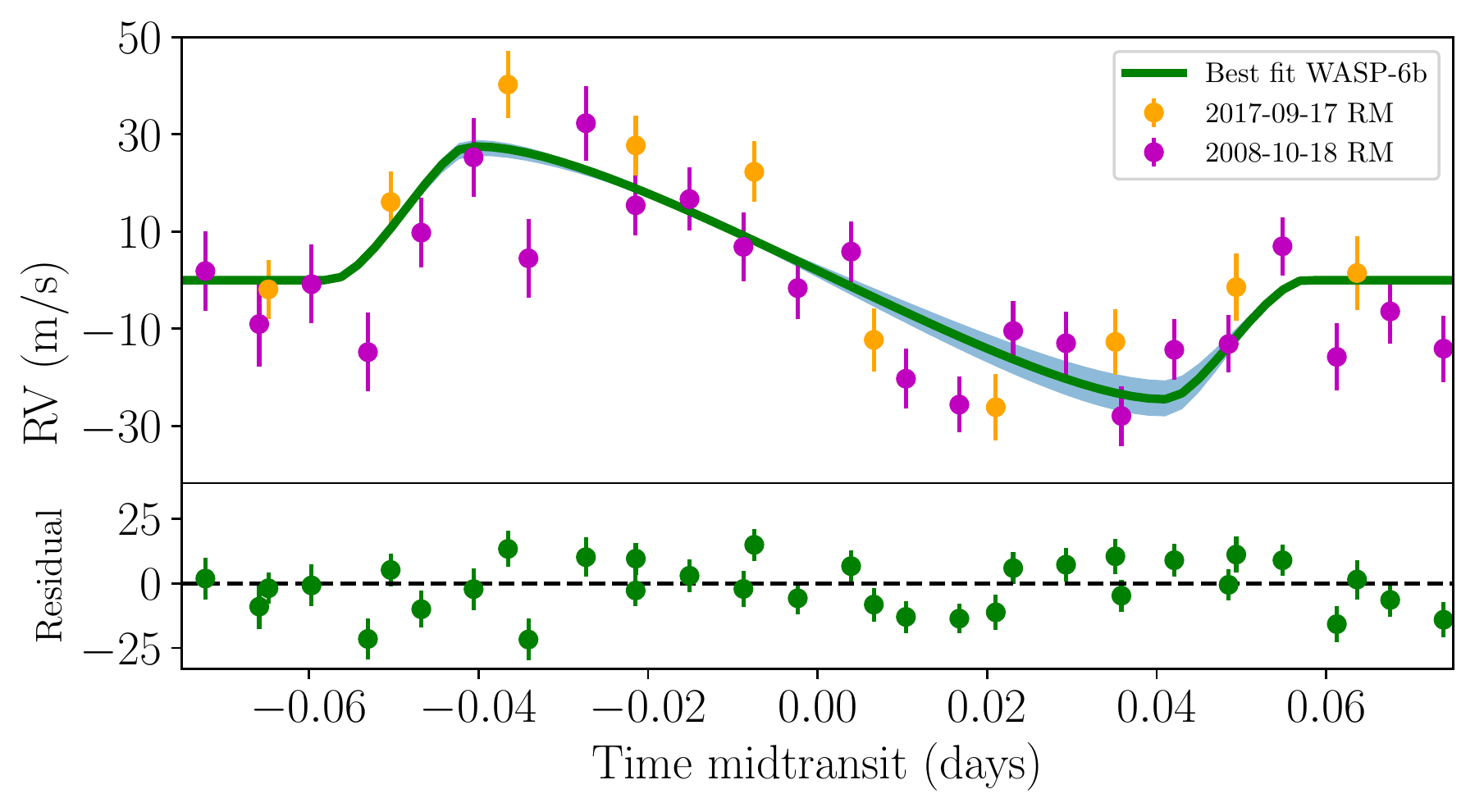}}
\medskip        \vspace{-1.cm} \hspace{0.cm} 
 \subfloat{\includegraphics[width=0.5 \textwidth]{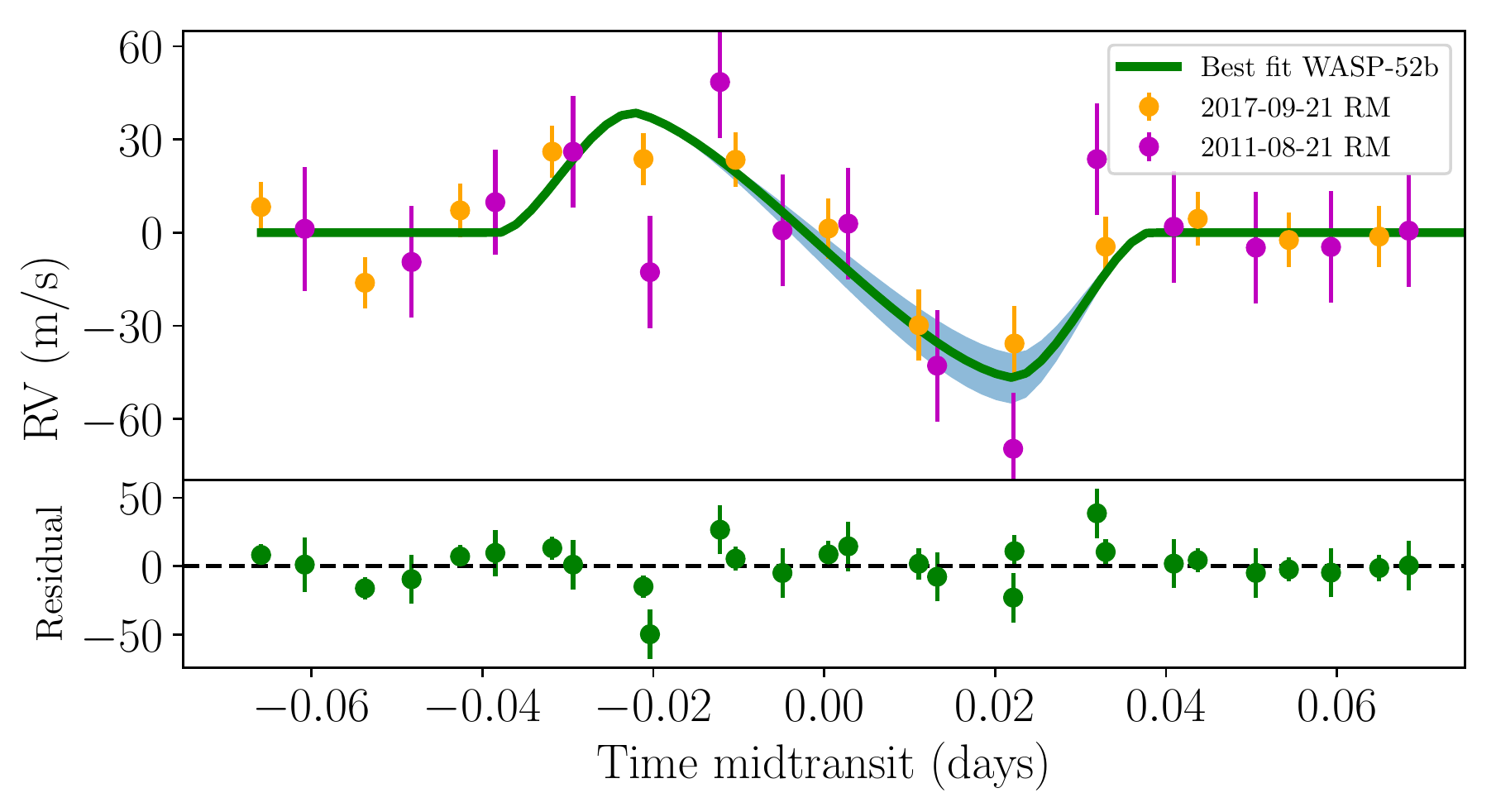}} \hspace{0.cm}\medskip
  \subfloat{\includegraphics[width=0.5 \textwidth]{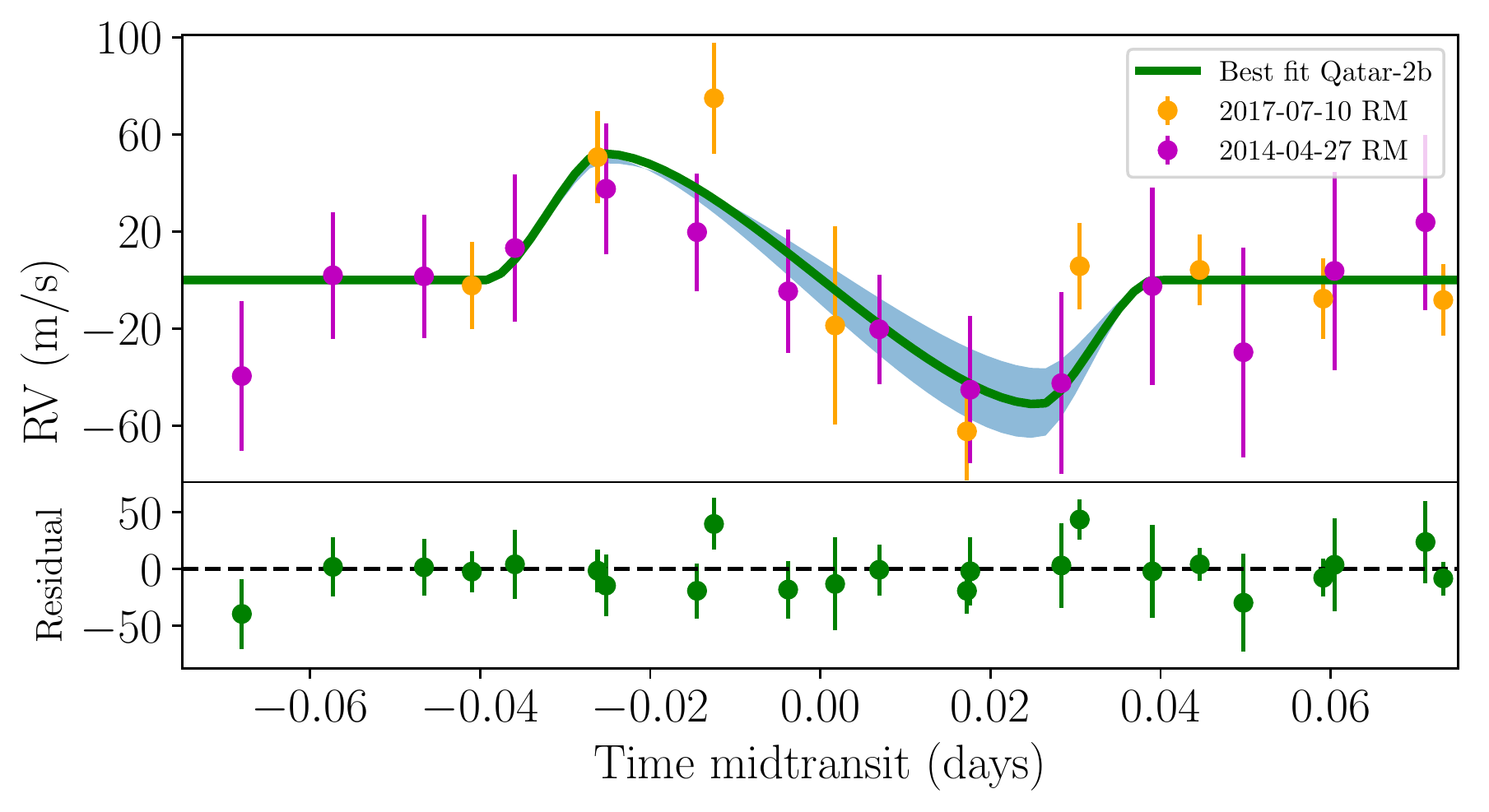}}

 \caption{Folded RM observations of our targets during several nights. The green line shows the best fitted RM model and blue region shows the zone where 68\% of the model solutions reside. In each panel the residuals are shown in the bottom panel.}

\end{figure*}

\begin{table*}
\caption{Details of the priors that we apply during our MCMC analysis}              \tiny

\centering                                      
\begin{tabular}{p{1.4cm} p{2.4cm} p{2.4cm} p{2.4cm} p{2.4cm} p{2.4cm} p{2.4cm}  }          
\hline\hline                        
Parameter & WASP-6b & WASP-19b & WASP-41b & WASP-52b & CoRoT-2b & Qatar-2b \\
\hline 
$\lambda (^\circ)$  &$\mathcal{U}(-180; +180)$&  $\mathcal{U}(-180; +180)$&  $\mathcal{U}(-180; +180)$&  $\mathcal{U}(-180; +180)$&  $\mathcal{U}(-180; +180)$ & $\mathcal{U}(-180; +180)$\\

$v \sin i (kms^{-1})$ & $\mathcal{N}(1.6; 0.6)$ & $\mathcal{N}(5.1; 0.5)$& $\mathcal{N}(2.66; 0.5)$ & $\mathcal{N}(3.6; 0.5)$ &  $\mathcal{N}(11.25; 4.)$& $\mathcal{N}(2.09; 0.5)$ \\

$T_{0} (JD)$ & $\mathcal{N}(8014.57580; 0.01)$ &  $\mathcal{N}(7878.63066; 0.01)$ & $\mathcal{N}(7858.64562; 0.01)$ & $\mathcal{N}(8017.65153; 0.01)$ & $\mathcal{N}(7946.62577; 0.01)$ &  $\mathcal{N}(7945.50129; 0.01)$\\
\hline                                             
\end{tabular}
\begin{flushleft} 
\textbf{Notes:}  $\mathcal{U}(a;b)$ is a uniform prior with lower and upper
limits of a and b; $\mathcal{N}(\mu; \sigma)$ is a normal distribution with mean $\mu$ and width $\sigma$. $T_{0}$ mean reported is the first transit time for each planet.

\end{flushleft}
\end{table*}

\begin{table*}
\caption{Best fitted values for $\lambda$ and $v \sin i$  obtained from each RM observations.}              
\centering                                      
\begin{tabular}{c c c c c c c c c}          
\hline\hline                        
Target & $\lambda$\#1 & $\lambda$\#2 & $\lambda$\#3 & $\lambda$\#Extra & $v \sin i$\#1 & $v \sin i$\#2 &  $v \sin i$\#3 &  $v \sin i$\#Extra \\
\hline
\\
WASP-6b & $-25.82^{+9.27}_{-10.26}$ & - & -  &  $5.32^{+11.53}_{-10.63}$& $1.34^{+0.15}_{-0.15}$ & - & - & $0.86^{+0.09}_{-0.08}$ \\
WASP-19b & $-5.55^{+3.74}_{-3.50}$ &  $-0.37^{+4.48}_{-4.17}$ & -  &  $10.38^{+3.20}_{-3.20}$& $3.78^{+0.33}_{-0.33}$ & $3.82^{+0.34}_{-0.35}$ & - & $2.93^{+0.25}_{-0.23}$ \\

WASP-41b & $5.47^{+4.45}_{-4.60}$ &  $19.39^{+4.01}_{-3.86}$ & -  &  $-1.72^{+6.00}_{-5.80}$& $1.38^{+0.07}_{-0.07}$ & $1.79^{+0.09}_{-0.08}$ & - & $1.42^{+0.09}_{-0.09}$ \\

WASP-52b & $5.38^{+5.48}_{-5.08}$ &  - & -  &  $12.03^{+8.79}_{-8.44}$& $1.78^{+0.20}_{-0.21}$ & - & - & $3.31^{+0.41}_{-0.41}$ \\

CoRoT-2b & $-4.22^{+6.65}_{-6.94}$ &  $-34.44^{+6.04}_{-6.28}$ & $-30.47^{+12.28}_{-13.88}$  &  $8.05^{+4.06}_{-3.73}$ & $7.29^{+0.44}_{-0.44}$ & $7.17^{+0.51}_{-0.53}$ & $6.48^{+0.91}_{-1.01}$ & $6.21^{+0.22}_{-0.22}$ \\

Qatar-2b & $-18.56^{+17.19}_{-18.42}$ &  - & -  &  $19.28^{+19.72}_{-19.10}$& $1.92^{+0.31}_{-0.33}$ & - & - & $1.78^{+0.38}_{-0.39}$ \\

\hline

\hline                                             
\end{tabular}
\begin{flushleft} 
 
\textbf{Notes:} $^a$ Obtained from recurring starspot occultations in photometric transit observations, $^b$ \citep{Tregloan-Reed-15}, $^c$ \citep{Tregloan-Reed-13}, $^d$ \citep{Southworth-16}, $^e$ \citep{Mancini-17}, $^f$ \citep{Nutzman-11}, $^g$ \citep{Mocnik-17}
\end{flushleft}
\end{table*}

\begin{table*}
\caption{Best fitted values for $\lambda$ and $v \sin i$  obtained from folded RM observations. }              
\centering                                      
\begin{tabular}{c c c c c c c  }          
\hline\hline                        
Parameter & WASP-6b & WASP-19b & WASP-41b & WASP-52b & CoRoT-2b & Qatar-2b \\
\hline 
$\lambda (^\circ)$  & $-9.49 ^{+7.05}_{-7.71}$ & $4.65 ^{+2.65}_{-2.51}$ & $9.15^{+2.85}_{-2.65}$& $5.47^{+4.61}_{-4.21}$& $6.01^{+3.22}_{-3.05}$& $−0.97^{+13.41}_{-13.03}$\\

$v \sin i (kms^{-1})$ & $0.98^{+0.08}_{-0.08}$  & $2.81^{+0.21}_{-0.21}$& $1.50^{+0.05}_{-0.05}$& $1.77^{+0.19}_{
-0.20}$& $6.28^{+0.18}_{-0.18}$& $1.79^{+0.27}_{-0.28}$ \\

Starspot crossing $\lambda (^\circ)$$^a$ & $7.2 \pm 3.7 $$^b$& $1.0 \pm 1.2$$^c$ &$6 \pm  11$$^d$ & $  3.8 \pm 8.4  $$^e$ & $  4.7  \pm  12.3   $$^f$ & $ 0  \pm  8  $ $^g$ \\

\hline                                             
\end{tabular}
\begin{flushleft} 
 
\textbf{Notes:} $^a$ Obtained from recurring starspot occultations in photomteric transit observations, $^b$ \citep{Tregloan-Reed-15}, $^c$ \citep{Tregloan-Reed-13}, $^d$ \citep{Southworth-16}, $^e$ \citep{Mancini-17}, $^f$ \citep{Nutzman-11}, $^g$ \citep{Mocnik-17}
\end{flushleft}
\end{table*}

\subsection{Fitting RM observation}
As the first step of our analysis we carry out a least-squares linear fit
to the out-of-transit RV data of each observed RM observations for each planet, and remove the linear trend. The main purpose of this is to eliminate the RV contribution from the Keplerian orbit (which we assume to be linear for the period of observations)\footnote{Using the shortest period of our sample,  0.78 days, the transit duration of 1.9 hours corresponds to 10 \% of the total phase, which supports our assumption of a linear trend from the Keplerian orbit.}, the systematics, and also the stellar activity which induces RV slopes to the out-of-transit RV data points \citep{Boldt-18}. After the linear trend is removed, we fit individual RM observation with the \textit{ARoME} model.

In our fitting procedure we consider the spin-orbit angle ($\lambda$) , projected stellar rotational velocity ($v \sin i$), and the mid-transit time ($T_{0}$) as our free parameters. Due to presence of degeneracy between $\lambda$ and $v \sin i$, especially for a low-impact parameter system \citep{Brown-17}, frequently these are the two common free parameters in fitting and analyzing RM signals. Since \textit{ARoME} generates an RM signal centered around zero time (assuming zero as the mid-transit), we have to remove the mid-transit times (calculated based on the reported ephemeris and orbital period of the planet) from our observation \footnote{https://exoplanetarchive.ipac.caltech.edu/cgi-bin/TransitView/nph-visibletbls?dataset=transits}. Due to the uncertainty on the reported value of the ephemeris and also the possible variation in transit time (due to the presence of an unknown companion in the systems) the calculated $T_{0}$ might not be very accurate; thus, we decided to leave $T_{0}$ as the third free parameter. It could be argued that the slope of the out-of-transit trend should be one of our free parameters in the fitting RM, but since \citet{Boldt-18} demonstrated that  trend removal does have a negligible impact on the spin-orbit angle $\lambda$ estimation, we decided to fit the slope and remove the trend prior to our RM fitting procedure. Nonetheless, to evaluate our choice, we explore the consequences of leaving the slope as an extra free parameter in our fitting procedure in Appendix A. The result of that test indicates that the impact is negligible on the other parameters, which support our choice.

The rest of the parameters required in \textit{ARoME}, such as stellar radius ($R_{\star}$), quadratic limb darkening coefficients of the star ($u_{1}$ and $u_{2}$), the semimajor axis of the planet, the orbital period of planet($a$), and the planetary orbital inclination angle ($i$), are fixed to their reported values in the literature (which are given in Table 1). We also fixed the macro-turbulence velocity of all stars to $\zeta= 4km s^{-1}$ and fixed the instrumental broadening according to the HARPS instrument profile ($\beta_{0}=1.3 km s^{-1}$). We would like to note that the macro-turbulence might not be an accurate estimate, but since we are only interested in the variation in $\lambda$, the inaccurate macro-turbulence will be the same on all our measured $\lambda$.  We also evaluate having macro-turbulence as an extra free parameter in our fitting procedure in Appendix B. Its results also demonstrate that the impact is negligible on the other parameters, which supports our choice.We also fixed the width of the
Gaussian ($ \sigma_{0}$) to the width of a Gaussian fit to the CCF of out-of-transit for each star. We note that we have two strong arguments for not letting these parameters  be free in our fitting procedure. The first  is that we want to have very similar fitting procedures to most of the RM studies, which only use $\lambda$ and $v \sin i$ as free parameters and fix the rest of the parameters to the values obtained from photometric transit. The second reason is based on the small number of data points in each RM observation. Thus, if we consider all these parameters free we would have more free parameters than observations and we might end up with an overfitted model.

The  best fit parameters and associated uncertainties in our fitting procedure are derived using a Markov chain Monte Carlo (MCMC) analysis,
using the affine invariant ensemble sampler emcee \citep{Foreman-Mackey-13}. The prior on $v \sin i$ and $T_{0}$ are controlled by Gaussian
priors centered on the reported value in the literature and width according to the reported uncertainties, and the prior on $\lambda$ is also controlled by a uniform (uninformative) prior between $\pm 180^\circ$. We list the selected type of priors and ranges for our free parameters for each target in Table 3.

We randomly initiated the initial values for our free parameters for 30
MCMC chains inside the prior distributions. For each chain
we used a burn-in phase of 500 steps,
judging the chain to be converged, and then again sampled
the chains for 5000 steps. Thus, the results
concatenated to produce  150000 steps. We determined the best fitted values by calculating the median values of the posterior distributions for each parameters, based on the fact that the posterior distributions were Gaussian.  The $1 \sigma$ uncertainties were taken to be the value enclosed in the 68.3 \% of the posterior distributions.

\subsection{Significant variation of spin-orbit angle}
Figure 1 shows the posterior distributions in $v \sin i - \lambda$ parameter space of all our targets delivered by the fit to individual RM obtained during different transits on different nights. We listed the best fitted values of $\lambda$ and  $v \sin i$ from individual RM observations in Table 4. We found that the estimated spin-orbit angle $\lambda$ of an exoplanet can be significantly altered (up to $42^\circ$) from transit to transit due to variation in the configurations of the stellar active regions  on different nights (mainly as a consequence of the stellar rotation and also the evolution of the stellar active regions)\footnote{There could be reasons other  than stellar activity that produce these variations, which we will discuss  in Sect. 7.}. The estimated $\lambda$ variation was larger than the simulation's result we described in a previous work, which suggested a variation of up to $15^\circ$ for hot Jupiters \citep{Oshagh-16}. The uncertainty on the estimated $\lambda$ also varies significantly from transit to transit. 

The estimated $v \sin i$ from the \textit{ARoME} fit, as mentioned in \citet{Brown-17}, are usually underestimated in comparison to their literature values; however, our goal here is not to measure their value accurately but to evaluate their variation from transit to transit. Our results depict a deviation of estimated $v \sin i$ from transit to transit; however, for most of the cases the variations are in the uncertainty ranges, and thus could be considered insignificant. However, we would like to note again that since \textit{ARoME} generally underestimates $v \sin i$ this result should be taken with a grain of salt.

The fitted transit time $T_{0}$ for all of our targets coincided with the respective calculated value (based on assuming an unperturbed Keplerian orbit with fixed periodic planetary orbit and according to reported ephemeris). Therefore, we conclude that our observations do not exhibit any signs of transit time variation for any of our targets. However, we would like to note that there might still be transit timing variation in our targets, but we could not detect it due to the small number of points in our RM observations, which prevented us from achieving high-precision transit time estimations.

We present each individual RM observation (obtained during different nights) for each of our targets with their best fitted model in Figures C.1--C.6 (Appendix C).

\section{Probing the correlations between $log R'_{HK}$ and spin-orbit angle variation}

In this section we  assess the presence of any meaningful
correlation between the amplitude of the variation of spin-orbit angle (largest variation $\Delta \lambda$) and measured stellar activity indicator $log R'_{HK}$\footnote{$log R'_{HK}$ is one of the most powerful chromospheric activity indicators and is estimated by measuring the excess flux in the core of
Ca ii H+K lines, normalized to the bolometric flux.} . The values of $log R'_{HK}$s were delivered as a by-product of DRS (as described in Lovis et al. 2011). We calculated the mean of measured $log R'_{HK}$s for our stars during our observation, and used the mean value as the measured $log R'_{HK}$. We also measured the standard deviation of $log R'_{HK}$s and used it as the uncertainty on the measured $\sigma_{log R'_{HK}}$. The value of $\sigma_{log R'_{HK}}$ by itself can also provide information about the variation in magnetic activity of stars during our observation, due to a stellar active region either appearing or disappearing   or to the occultation of the active region by a transiting planet.

We inspected the presence of a correlation based
on the Spearman's rank-order correlation coefficient ($\rho$)\footnote{Spearman's rank-order correlation assesses
how well two variables can be described with a monotonic relationship
that is not purely linear.}. We evaluated the significance of correlation
using the Bayesian approach described in \citet{Figueira-16}. The posterior distribution of $\rho$ indicates
 the range of $\rho$ values that is compatible with
the observations. 

Figure 2, top panel, presents the correlations between  $\Delta \lambda$ and $log R'_{HK}$.  We note that Qatar-2, due to its faintness ($V=13.3$) and low signal-to-noise ratio in the region of spectrum in which $log R'_{HK}$ is estimated, has only one estimated $log R'_{HK}$; therefore, it was impossible to measure its  $\sigma_{log R'_{HK}}$.  As this figure shows, there is no meaningful correlation between the measured value of  $\Delta \lambda$ and $log R'_{HK}$. Figure 2, bottom panel, displays the correlations between  $\Delta \lambda$ and $\sigma_{log R'_{HK}}$, which show an anti-correlation but not a significant correlation. We note that in this plot we discarded Qatar-2.

\section{Folding several RM observations}
The most logical approach for eliminating the effect of stellar activity on RM observations, and thus minimizing its impact on the estimated spin-orbit angle $\lambda$, is to obtain several RM observations and combine and fold them. This is based on the fact that the configuration of the stellar active region evolves from transit to transit, but the planetary RM signal remains constant; thus, averaging several RM observations will average out the activity noise and increase the planetary RM signal-to-noise ratio. Therefore, in this section we fold all the observed RM observations for each planet (analyzed in the previous section), and try to repeat the fitting procedure to estimate more accurately the spin-orbit angle $\lambda$ and the host star $v \sin i$ .

In order to fold all the available RM observations for each target, we utilize the best fitted transit time ($T_{0}$) obtained in the previous section. Therefore, in our fitting procedure we  no longer leave the transit time as a free parameter, thus  $\lambda$ and $v \sin i $ are our only free parameters.

Figure 3 presents the posterior distributions from the fit to folded RM observations of all of our targets, and displays the best fitted values for $\lambda$ and $v \sin i $ and their associated  $1 \sigma$ uncertainties. We summarized the best fitted values in Table 5. Moreover, in Table 5 we also list the estimated values of $\lambda$ for our targets, which were estimated from an independent method of recurring starspot occultations in photometric transit light curve observations. The comparison between our estimated $\lambda$ from folded RM observations and those measured from recurring starspot occultations (both presented in Table 5) reinforce that the strategy of folding RM observations adequately eliminates the stellar activity effect, and provides an accurate estimate of $\lambda$. We present the folded RM observations for each of our targets and their best fitted model in Figure 4.

\begin{table*}[h!]
\caption{Standard deviation of the residual between SOAP3.0 predicted RM and observed RM for models that ignore the spot crossing events and for models that take into account spot crossing events.}              
\centering                                      
\begin{tabular}{c c c c c c  }          
\hline\hline                        
Standard deviation ($ms^{-1}$)& CoRoT-2b\#1 & CoRoT-2b\#2 & WASP-19b\#1 & WASP-19b\#2 & Qatar-2b  \\
\hline 
No spot &33.862& 82.58&18.75&18.22 & 23.09\\
With spot & 27.93&65.77& 16.50& 17.23&18.56\\

\hline                                             
\end{tabular}
\begin{flushleft} 
 
\end{flushleft}
\end{table*}

\section{Simultaneous photometric transit light curve and RM observations}

Photometric transit observations can usually be acquired on much shorter exposure times than RM observations\footnote{Spectrographs lose photons due to slit losses,  stray light, and scattered light. Moreover, spectrographs disperse photons over the detector where each pixel has a different readout noise. Therefore, spectrograph by construction required many more photons to reach  the same S/N as photometric observations. },  and thus could lead to the easier and clearer detection of the occultation of  active regions during the transit of an exoplanet. We obtained five simultaneous photometric transit observations with our RM observations, four with the TRAPPSIT telescope and one with SPECULOOS. All the photometric transit light curves clearly indicated the presence of  active regions' crossing. Thus, the main aim of this section is to assess whether  we can have a better RM modeling by having the information from the active region's crossing event during a photometric transit (which provides information about the size, position, and contrast of active regions).

\subsection{Fitting photometric transit anomalies with SOAP3.0}
In this section we  use the publicly available tool \textit{SOAP3.0}. This tool has the capability of simulating a transiting planet and a rotating star covered with active regions, and delivers photometric and RV variation signals. \textit{SOAP3.0} takes into account not
only the flux contrast effect in these regions, but also the RV
shift due to inhibition of the convective blueshift inside these
regions. \textit{SOAP3.0} also takes into account the occultation between the transiting planet and active regions in its calculation of transit light curve and RM.

We use SOAP3.0 to obtain the best fitted model to the photometric transit light curves and then compare the corresponding RM of the best fitted model  with the observed RM observations. Because of the slowness of \textit{SOAP3.0}, due to its numerical nature in comparison to the analytical nature of \textit{ARoMe}, performing an MCMC approach using \textit{SOAP3.0} is not feasible. Therefore, we decided to perform a
$\chi^{2}_{reduced}$ minimization using \textit{SOAP3.0} to fit the active region crossing events in the photometric transit light curves. From visual inspection we could identify only three active region crossing events during each observed transit light curve, thus we decided to fix the number of active regions to three.  In $\chi^{2}_{reduced}$ minimization we fixed all the required parameters of stars and planets in \textit{SOAP3.0} (the same  parameters in Table 1) except for the three active regions' parameters (filling factor, location, and temperature contrast) and let them vary as free parameters. The range of free parameters which were explored are listed as the spots' filling factor [0.1\%:20\%], latitude [$-90^\circ$:$+90^\circ$], longitude [$0^\circ$:$360^\circ$], and temperature contrast [0:$-T_{eff}$].

\subsection{Improved RM prediction}
As mentioned before, here we intend to compare the corresponding RM of the best fitted model to the transit light curves with the observed RM \footnote{We did not perform SOAP3.0 fitting to the RM observation, and we only fitted SOAP3.0 to the photometric transit-light curve and used the corresponding RM of the best fitted model.}. We overplotted the RM counterpart of the best fitted model over the simultaneous observed RM observations for WASP-19b, CoRoT-2b, and Qatar-2b in Figures 5, 6, and 7, respectively. This result demonstrated that the RM counterparts coincide much better with the observed RM. As a consequence, the residual between the \textit{SOAP3.0} RM whose considered active region crossing is lower than the residual of \textit{SOAP3.0} RM without any spot crossing event. To quantify the improvement, we computed the standard deviation of the residual between SOAP3.0 best fit predicted RM and the observed RM, for models ignoring  stellar spots and for models taking them into account. All standard deviations are listed in Table 6, and they all support the improvement in the RM prediction.

It is worth mentioning that a strong degeneracy exists between the parameters of the stellar active regions; for instance, different filling factor, position, and contrast could produce very similar photometric signatures while generating completely different RV signals. Therefore, we have to clarify our best fitted model to the photometric transit light curves, and also the best fitted value for the active regions' parameters might not correspond  to  accurate values, although they more closely predict the RM observation than considering no active region occultation.

\begin{figure}[h!]
\includegraphics[width=0.45 \textwidth, height=10cm]{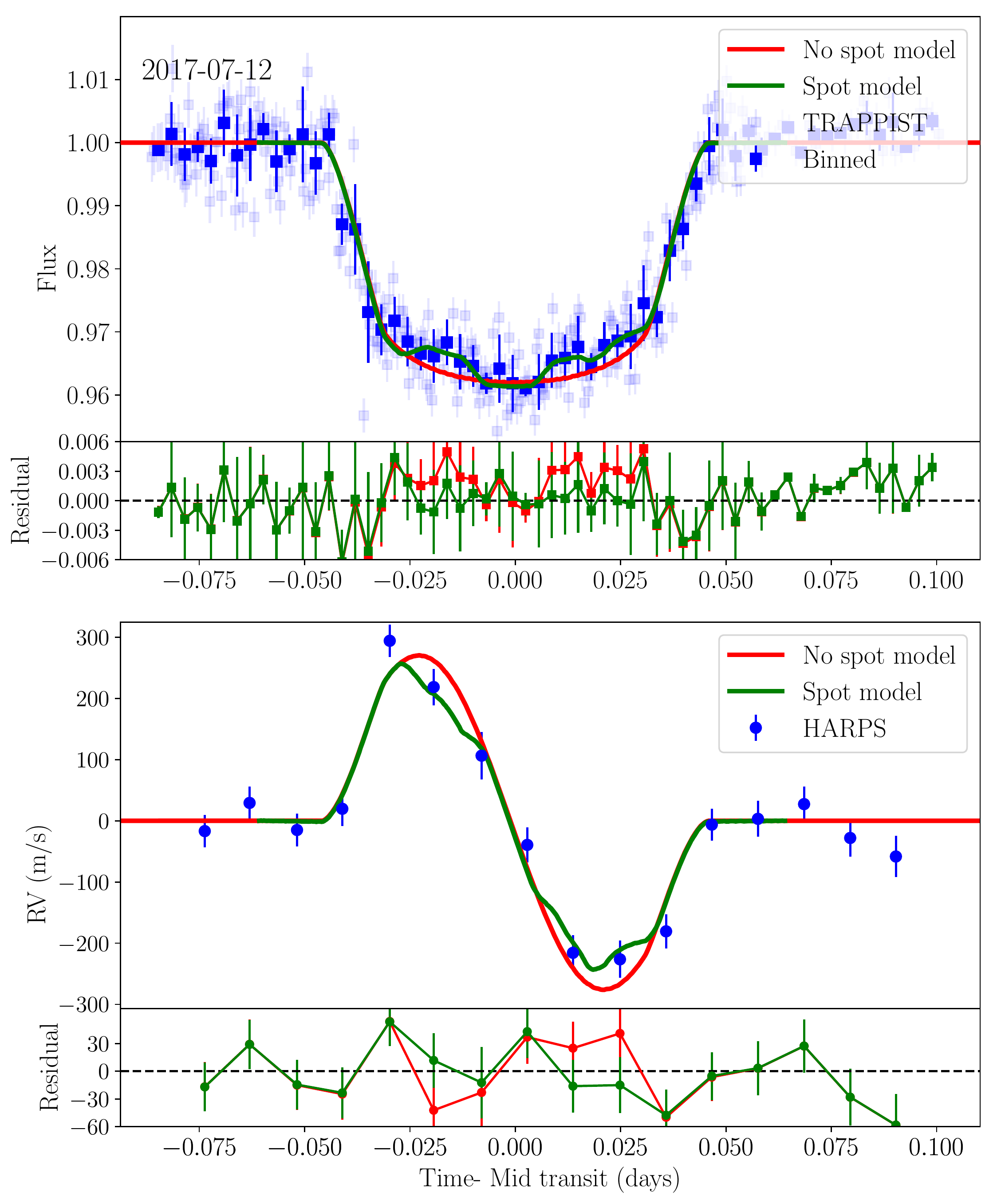}

\includegraphics[width=0.45 \textwidth, height=10cm]{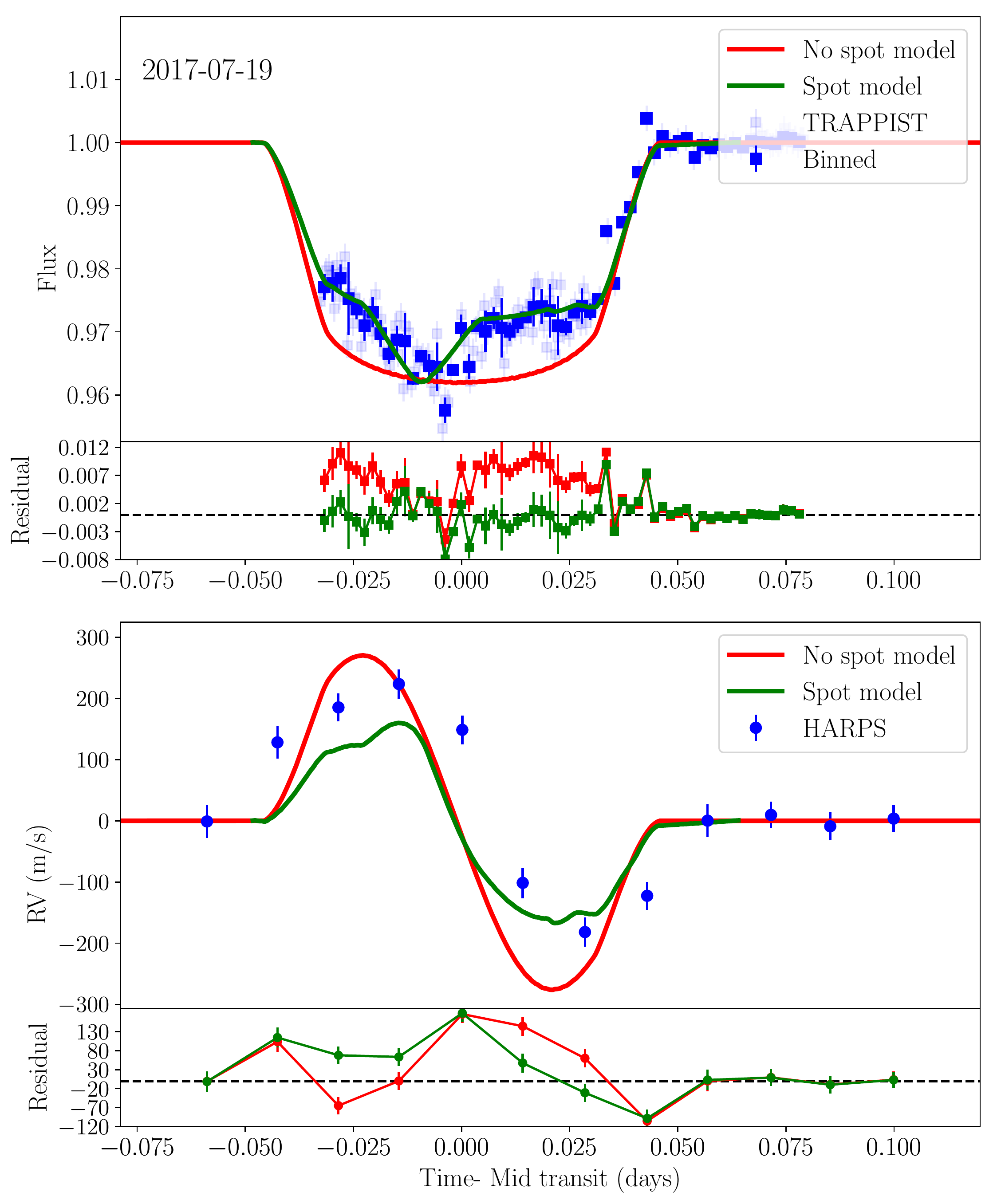}
\caption{Simultaneous photometric transit and RM observation of CoRoT-2b on the night of 2017-07-12 (top panels) and on night the night of 2017-07-19 (bottom panels). The dark blue square represents the binned TRAPPIST photometric observations, and the dark blue filled circle the HARPS RM observation. The red line is the SOAP3.0 model without considering any stellar active regions. The green lines are the SOAP3.0 best fitted model to transit light curve taking into account three spots.}
\end{figure}

\begin{figure}[ht]
\includegraphics[width=0.45 \textwidth, height=10cm]{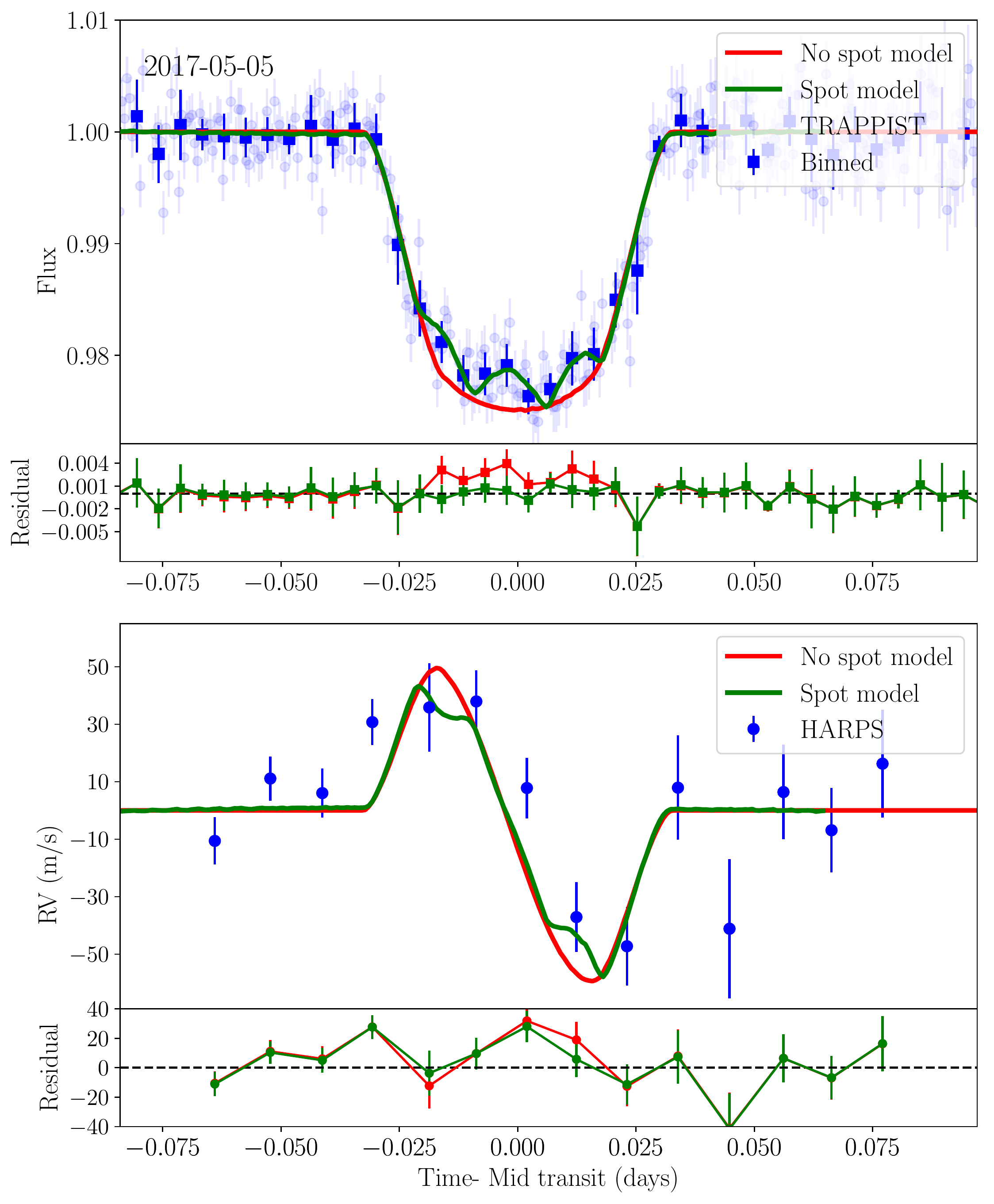}

\includegraphics[width=0.45 \textwidth, height=10cm]{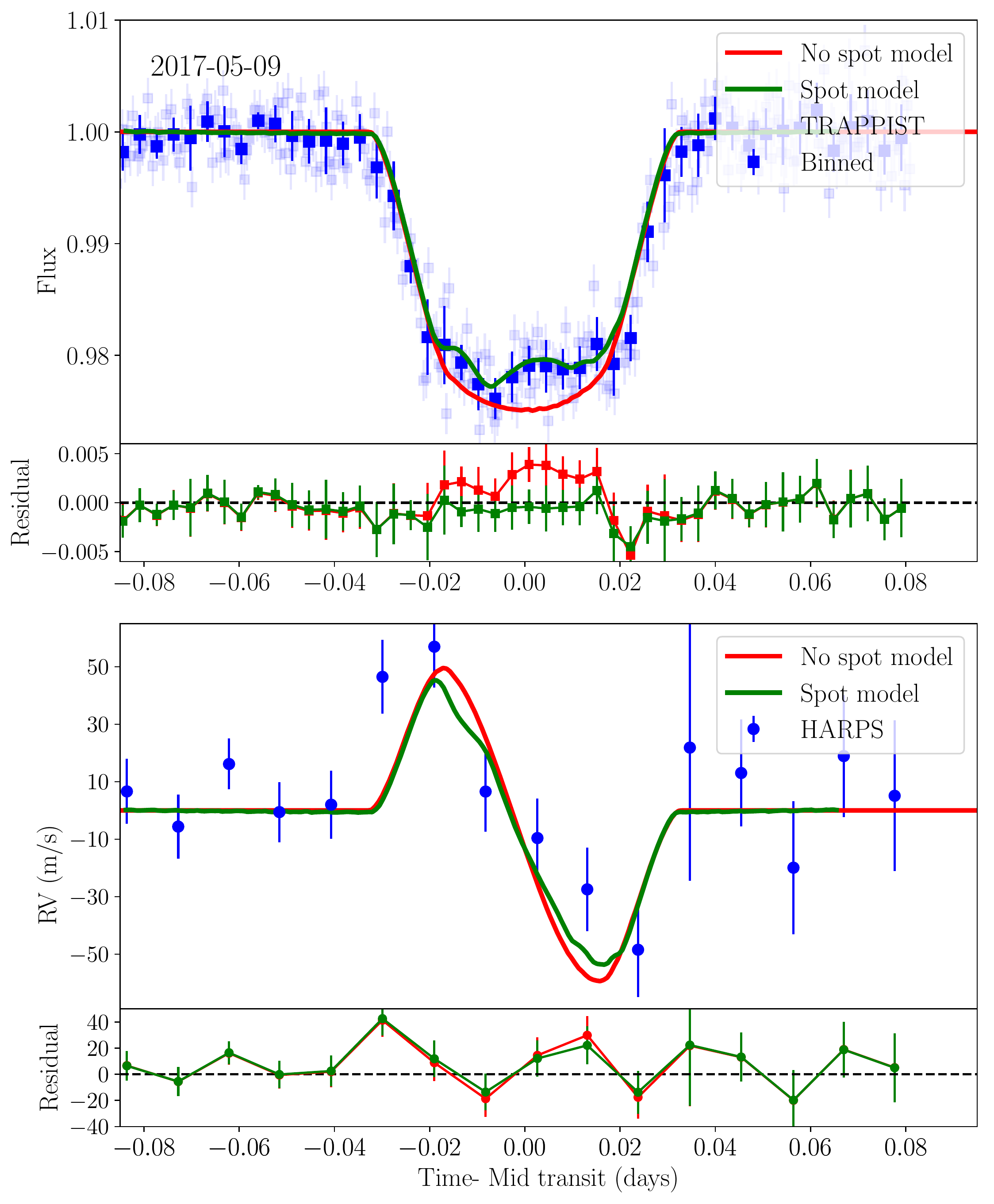}
\caption{Simultaneous photometric transits and RM observations of WASP-19b on the  night of 2017-05-05 (top panels) and on the night of 2017-05-09 (bottom panels). The lines and points are the same as in Fig. 3.}
\end{figure}

\begin{figure}[ht]
\includegraphics[width=0.45 \textwidth, height=10cm]{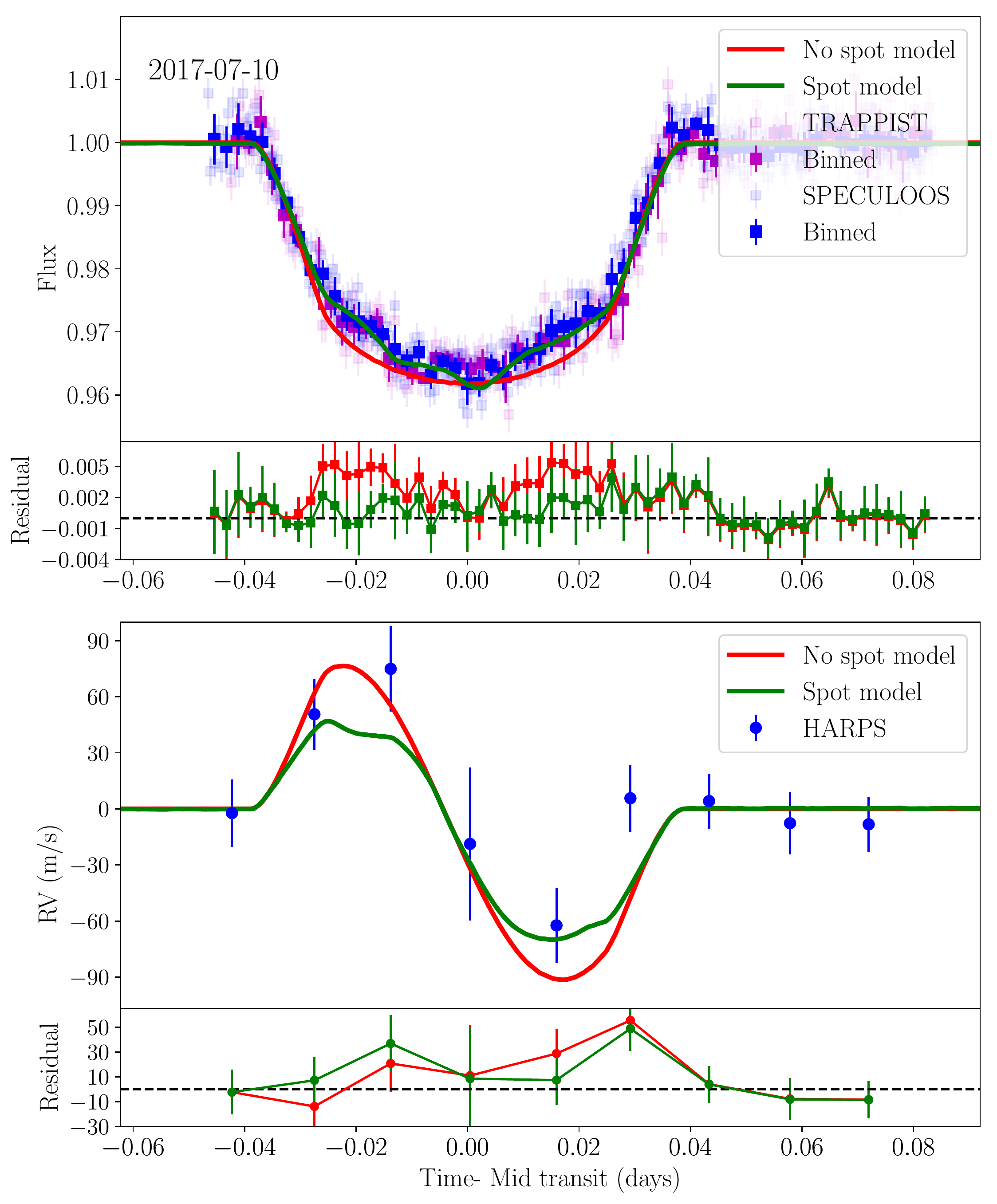}
\caption{Simultaneous photometric transits and RM observations of Qatar-2b on the night of  2017-07-10. The photometric observations were obtained from simultaneous observations of TRAPPIST and SPECULOOS. The lines and points are the same as in Fig. 3.}
\end{figure}

\section{Discussion}
Estimation of $\lambda$ and $v \sin i$ could be influenced by second-order effects such as the stellar convective blueshift and granulation \citep{Shporer-11, Cegla-16}, the microlensing effect due to the transiting planet's mass \cite{Oshagh-13c}, the impact of ringed exoplanet on
RM signal \citep{Ohta-09, deMooij-17, Akinsanmi-18}, and stellar differential rotation \citep{Albrecht-12, Cegla-16b, Serrano-18}. However, their expected
signals in RM observations are different from the active region crossing events. More importantly, all of their signals are constant during several transits; thus, even if they affect the estimation of $\lambda$ and $v \sin i$, their influence cannot produce variation in the estimated $\lambda$. Therefore, of the extensive list of effects above, we can conclude that the variation in the estimated $\lambda$ could have only  originated from the stellar activity noise.

The simulations presented in \citet{Oshagh-16} predicted that the variation in  estimated $\lambda$ could reach up to $15^\circ$ for hot Jupiters; however, our observational campaign shows a variation that is twice as large.  The plausible explanation for this underestimation of variation in the simulation could be that in the simulation the stellar active regions were considered to be similar to the sunspots (e.g., a filling factor of around 1\%). However, all the stars in our sample exhibit a much higher level of activity than the Sun, and are covered with much larger stellar spots (filling of stellar spots on the WASP-6, WASP-19, WASP-41, WASP-52, CoRoT-2, and Qatar-2 were 6\%, 8\% , 3\%, 15\%, 16\%, and 4\%, respectively).

We would like to note that throughout one RM observation the target's airmass varies, and also from one night to another  the mid-transit occurs at different airmass. Moreover, the seeing condition fluctuates from night to  night. Therefore, there could be some considerable contribution from airmass and seeing
variations which might lead to the variation in observed RM observations. Although this statement is provable, correcting their effect is not a trivial task and beyond the scope of the current study. However, this again points to the fact that having only single-epoch RM observation could be vulnerable to other unaccounted noise sources. Moreover, we would like to suggest obtaining several RM observations of a planet transiting a very inactive star to be able to better explore  the seeing and airmass conditions on the estimated $\lambda$.

We showed that folding several RM observation could mitigate the impact of stellar active region occultation; however, if RM observation are done on consecutive transits, and are not separated with long time interval from each other, we would like to note that they can be affected by occultation with the same active regions, and thus folding the RM observations will not improve the accuracy of the estimation of $\lambda$. Therefore, we suggest  obtaining several RM observations with a long time separation compatible with several stellar rotations to ensure that RM observations are affected by the configurations of different active regions.

Our results also highlighted the power of having simultaneous photometric transit observations with RM observations, which provides unique information about the stellar active regions that have been occulted during the transit, and leads to a better elimination of their influence on the observed RM. Thus, in the cases where only one RM observation can be observed (e.g., due to the long periodicity of the planet) and the combination of several RM observations is not feasible, having simultaneous photometric transit will be needed and crucial. Although, one missing piece in analyzing simultaneous photometric and RM observations is the lack of an analytical model, similar to \textit{ARoME} which takes into account the active region occulation in RM modeling, which reduces significantly the computational cost and allows a more robust fitting utilizing the MCMC approach.

Complementary methods such as
Doppler tomography have been used to estimate the spin-orbit angle $\lambda$ of a planet around hot and rapidly rotating host stars
which the conventional RM technique is unable to deal with. However, the impact of a stellar active region (either occulted and unocculted ones) on the Doppler tomography signal has not been explored, and by having our observation data set we will be able to explore this matter. However, probing this effect is beyond scope of the current paper and will be pursued in a forthcoming publication.

\citet{Oshagh-16} predicted that the impact of active regions' occultation on the estimated $\lambda$ will be more significant for the Neptune- or Earth-sized planets. Since the planets in our sample are all hot Jupiters (gas giant exoplanets orbiting very close to their host stars), we can extrapolate and speculate that accurately estimating the spin-orbit angle for a small-sized planet will be a challenging task. As an example of this difficulty and complication for small-sized planet we can point  to the case of 55 Cnc e whose different RM observations lead to different interpretations regarding its spin-orbit angles \citep{Bourrier-14, Lopez-Morales-14}. 

If the variations in $\lambda$ are mostly ascribed to the stellar activity, they should depend on the wavelength region where RVs are measured. Therefore, performing chromatic RM observations similar to \citet{DiGloria-15}, and measuring $\lambda$ variation in different wavelengths could provide information about which wavelength range the stellar activity influence is minimum and thus the estimated  $\lambda$ is more accurate. However, exploring this area is beyond the scope of the current paper and will be pursued in forthcoming publication.

\section{Conclusion}
Rossiter--McLaughlin observations have provided an efficient way to estimate spin-orbit angle $\lambda$ for more than 200 exoplanetary systems which include planets on
highly misaligned, polar, and even retrograde orbits. So far, however,  mostly single-epoch RM observations have been used to estimate the spin-orbit angle, and therefore there has been no study evaluating the dependence of estimated spin-orbit angle on induced noise in RM observations. One of the most important and dominant sources of time varying noise in RM observations is  stellar activity. In this paper we obtained several RM observations of known transiting planets which all transit extremely active stars, and by analyzing them individually we were able to quantify, for the first time, the variation in the estimated spin-orbit angle from transit to transit. Our results reveal that the estimated spin-orbit angle can be significantly altered (up to $\sim 42^\circ$). This finding is almost two times larger than the expected variation predicted from the simulation. We could not identify any meaningful correlation between the variation of estimated spin-orbit angles and stellar magnetic activity indicators. We also investigated two possible approaches for mitigating the influence of  stellar activity  on RM observations. The first strategy is based on obtaining several RM observations and folding them to reduce the stellar activity noise. Our results demonstrate that  this is a feasible and robust way to overcome this issue. The second approach is based on acquiring simultaneous high-precision short-cadence photometric transit light curve that can provide more information about the properties of the  stellar active region, which  will allow a better RM modeling.

\begin{acknowledgements}

\scriptsize M.O. acknowledges research funding from the Deutsche
Forschungsgemeinschft (DFG, German Research Foundation) - OS 508/1-1.  We acknowledge
the use of the software packages pyGTC \citep{Bocquet2016} and emcee \citep{Foreman-Mackey-13}. The research leading to these results has received funding from the European Research Council (ERC) under the FP/2007-2013 ERC grant agreement no. 336480 and from an Actions de Recherche Concert\'ee (ARC) grant, financed by the Wallonia-Brussels Federation (PI Gillon). This work was also partially supported by a grant from the Simons Foundation (PI Queloz, grant number 327127), and by the MERAC foundation (PI Triaud). L.D. acknowledges support from the Gruber Foundation Fellowship. M.G. and E.J. are F.R.S.-FNRS Research Associate and Senior Research Associate, respectively.  We would like to thank the referee, Teruyuki Hirano, for his constructive comments and insightful suggestions, which
added significantly to the clarity of this paper.

\end{acknowledgements}

\bibliographystyle{aa}
\bibliography{mahlibspot}

\begin{appendix}

\section{RV slope of out-of-transit as free parameter}
In this section we  evaluate the consequence of leaving the slope of out-of-transit RM as an extra free parameter in our fitting procedure.  We consider the linear trend with two parameters $m$ and $b$ ($m\times T+b$), and include them in our free parameters in the MCMC fitting procedure. In order to test this, we selected the case of CoRoT-2b which has the greatest number of RM observations, and repeated the fitting procedure on each individual RM observation of CoRoT-2b. We considered a uniform (uninformative) prior on both $m$ and $b$. We present the posterior probability distributions of all four free parameters in Figure A.1. As our results show, the posterior  distributions in $v \sin i - \lambda$ parameter space is not affected by leaving the slope as a free parameter (cf. the posterior distribution of  $v \sin i - \lambda$  in this figure with the posterior distribution of  $v \sin i - \lambda$  in Figure 1). Therefore, our test demonstrates that the impact of leaving the out-of-transit's slope on the estimated $\lambda$ is negligible, which supports our choice, and is also in agreement with result of \citet{Boldt-18}. We also present the best fitted model (having slope as a free parameter) and the individual RM observations of CoRoT-2b in Figure A.2. 

\begin{figure}
\includegraphics[width=0.45 \textwidth, height=8cm]{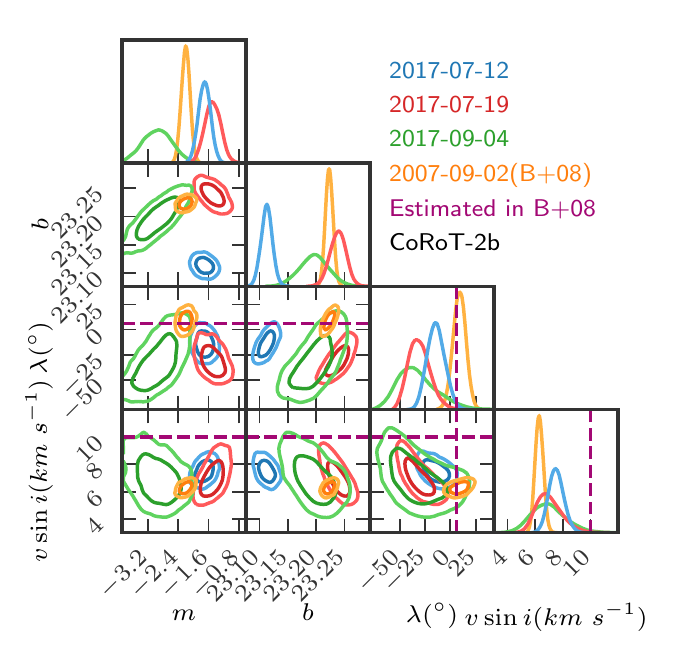}
\caption{Posterior probability distributions for four free parameters (including the slope of out-of-transit) space of fit to individual RM observations of CoRoT-2b.}
\end{figure}

\begin{figure}
\includegraphics[width=0.45 \textwidth, height=12cm]{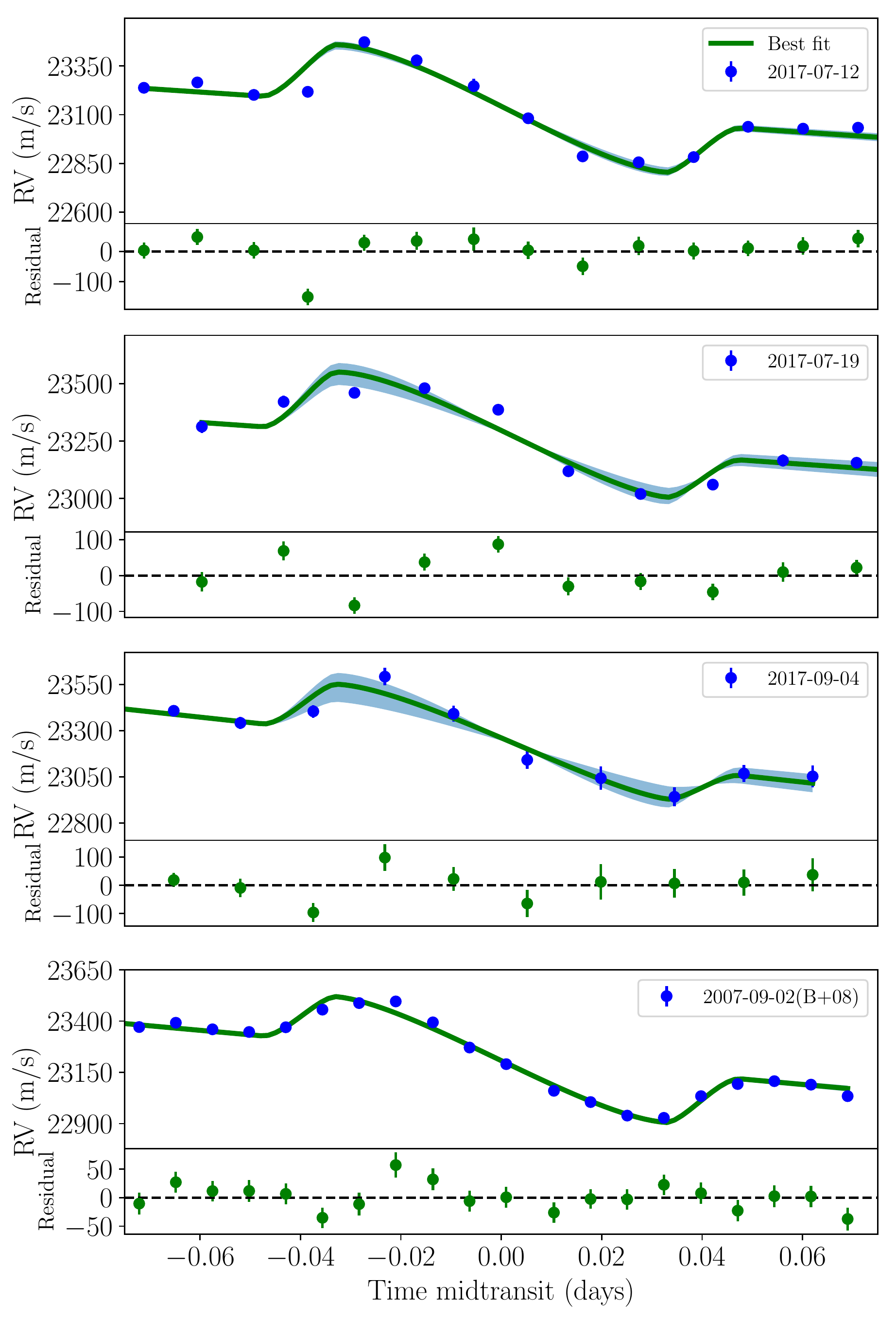}
\caption{RM observations of CoRoT-2b during several nights. The green line displays the best fitted RM model (also considering the slope of out-of-transit  as free parameter). The blue region shows the zone where 68\% of the model solutions reside. In each panel the residuals are shown in the bottom panel.}
\end{figure}

\section{Macro-turbulence as free parameter}

In this section we evaluate the consequence of leaving the macro-turbulence ($Z$) as an extra free parameter in our fitting procedure. Similar to Appendix A, we tested it on CoRoT-2b RM observations and repeated the fitting procedure on each individual RM observation.  We considered a uniform (uninformative) prior on $Z$ from 1 to 10 $kms^{-1}$. We present the posterior probability distributions of all three free parameters in Figure B.1. As our results show, the posterior  distributions in $v \sin i - \lambda$ parameter space is not affected by leaving macro-turbulence as a free parameter (cf. the posterior distribution of  $v \sin i - \lambda$  in this figure with the posterior distribution of  $v \sin i - \lambda$  in Figure 1). Therefore, our test shows that the impact of leaving macro-turbulence as a free parameter does not affect the estimated $\lambda$, which supports our choice.

\begin{figure}
\includegraphics[width=0.45 \textwidth, height=8cm]{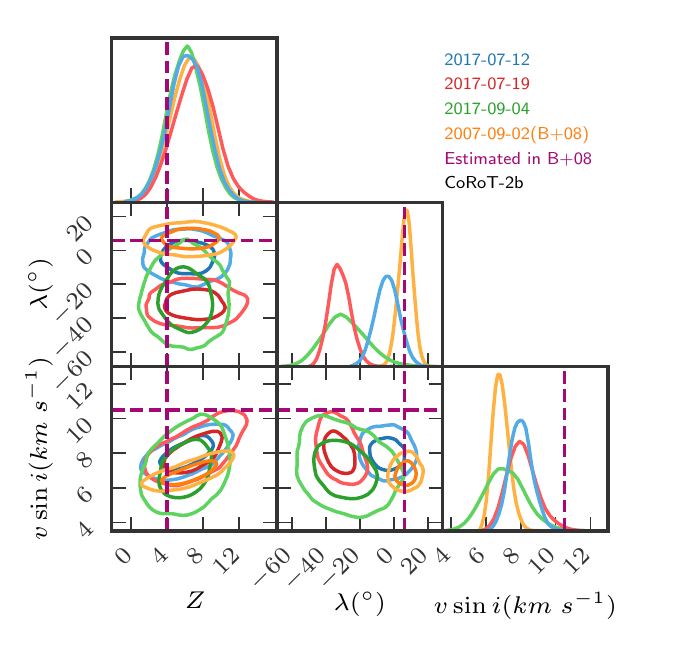}
\caption{Posterior probability distributions of three free parameters (including  macro-turbulence) space of fit to individual RM observations of CoRoT-2b. }
\end{figure}

\section{Best fitted model to individual RM observations}
In this section we  present each individual RM observation (obtained during different nights) for each of our targets with their best fitted model, as obtained in Sect 3. Moreover, we overplotted the best fitted model to the folded RM (dashed red line) to help the readers to visually identify the variation in the RM curves from transit to transit.

\begin{figure}
\includegraphics[width=0.45 \textwidth, height=12cm]{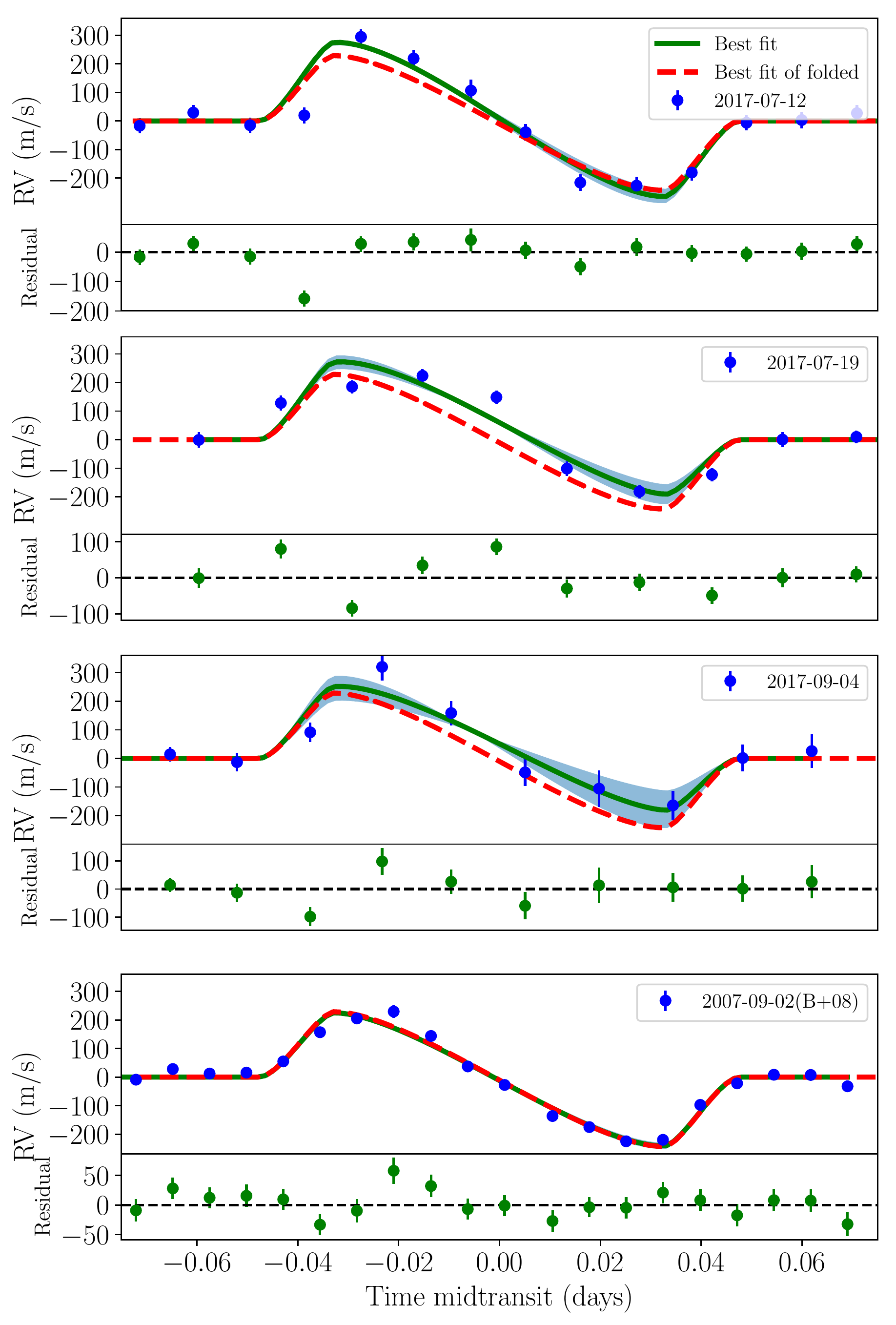}
\caption{RM observation of CoRoT-2b during several nights. The green line displays the best fitted RM model and the blue region shows the zone where 68\% of the model solutions reside. In each panel the residuals are shown in the bottom panel. The red dashed lines show the best fitted RM model obtained from the folded RM observations from Figure 4.}
\end{figure}

\begin{figure}
\includegraphics[width=0.45 \textwidth, height=12cm]{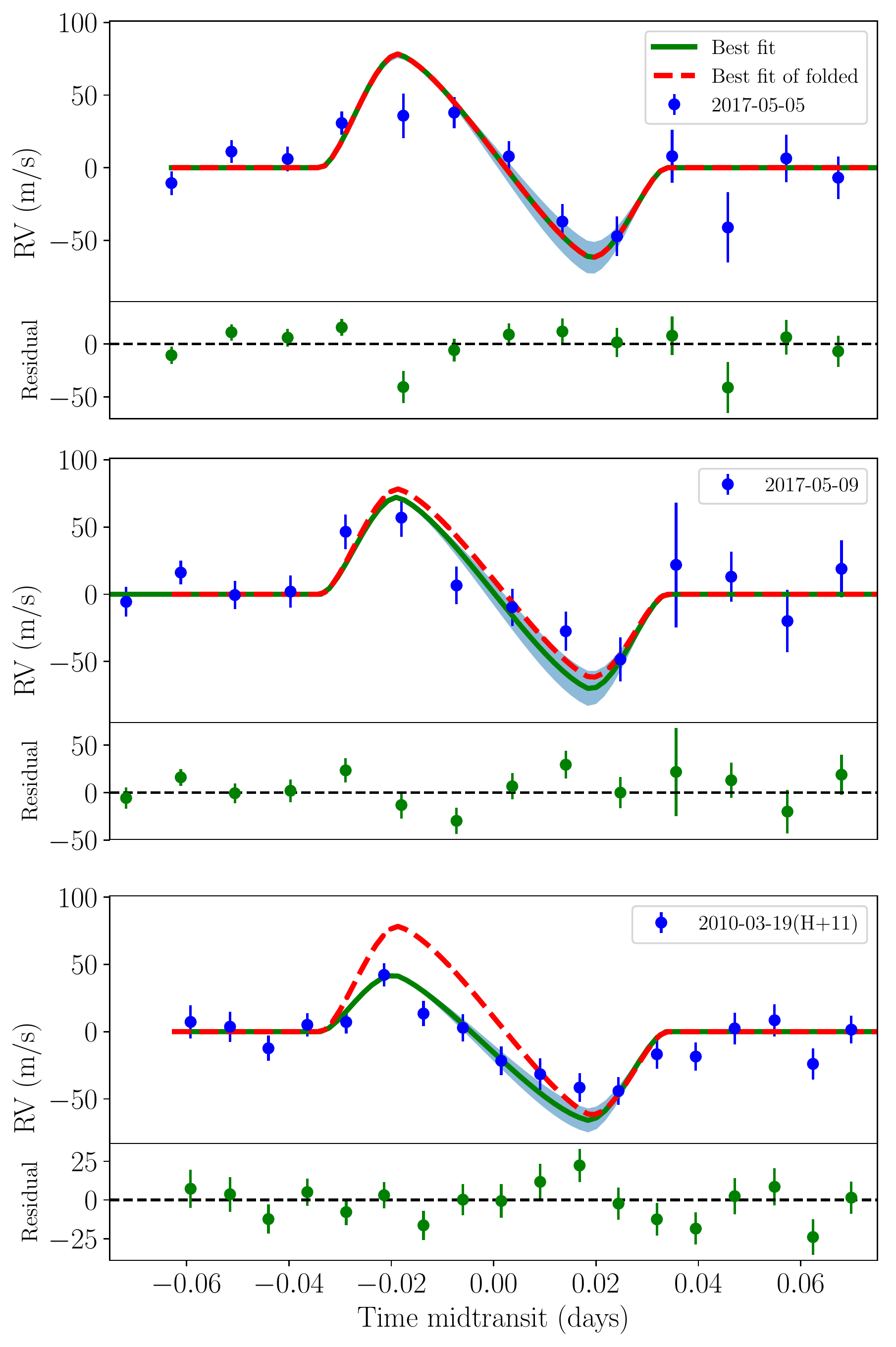}
\caption{As for Fig. C.1, but for WASP-19b. }
\end{figure}

\begin{figure}
\includegraphics[width=0.45 \textwidth, height=12cm]{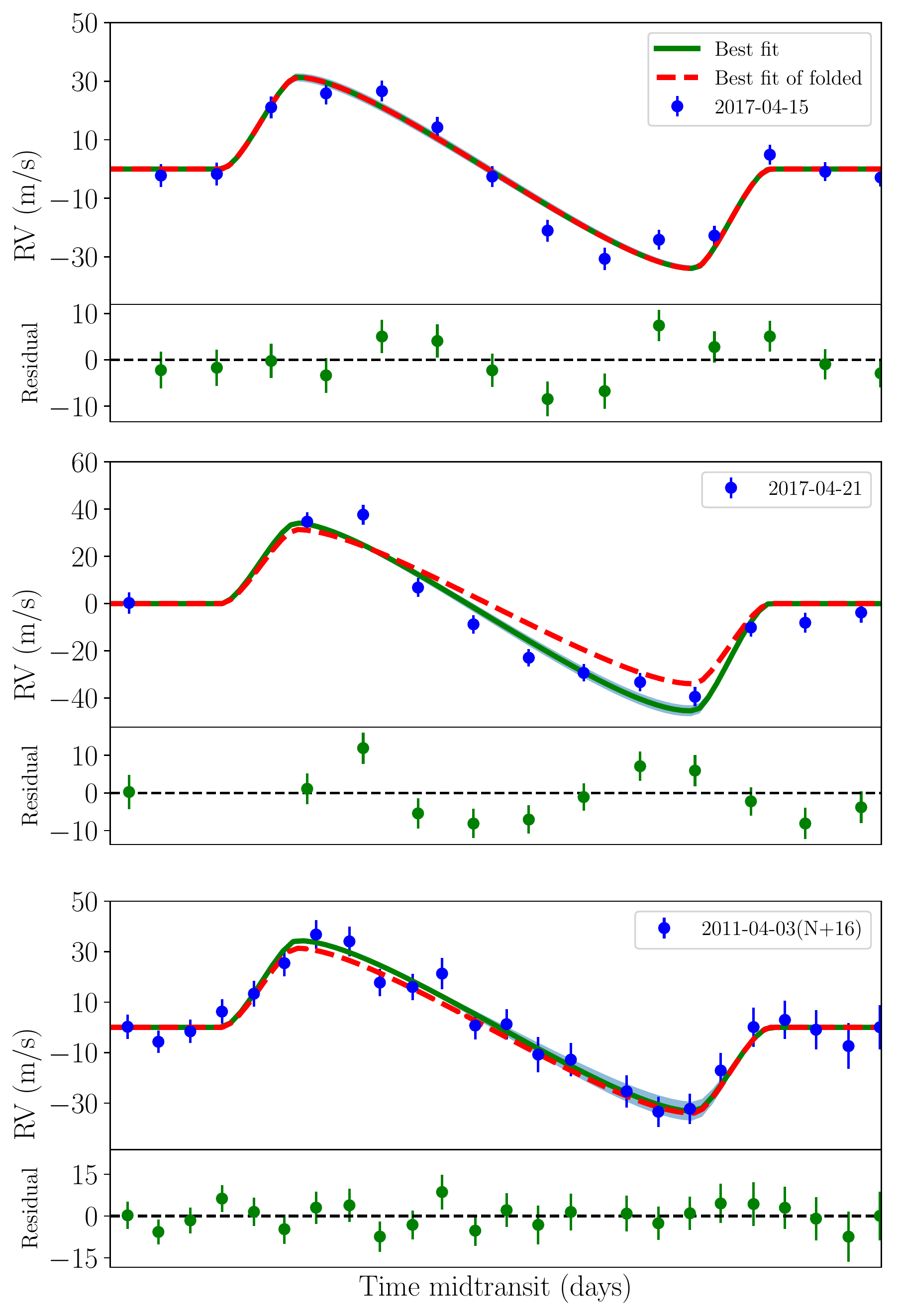}
\caption{As for Fig. C.1, but  for WASP-41b. }
\end{figure}

\begin{figure}
\includegraphics[width=0.45 \textwidth, height=8cm]{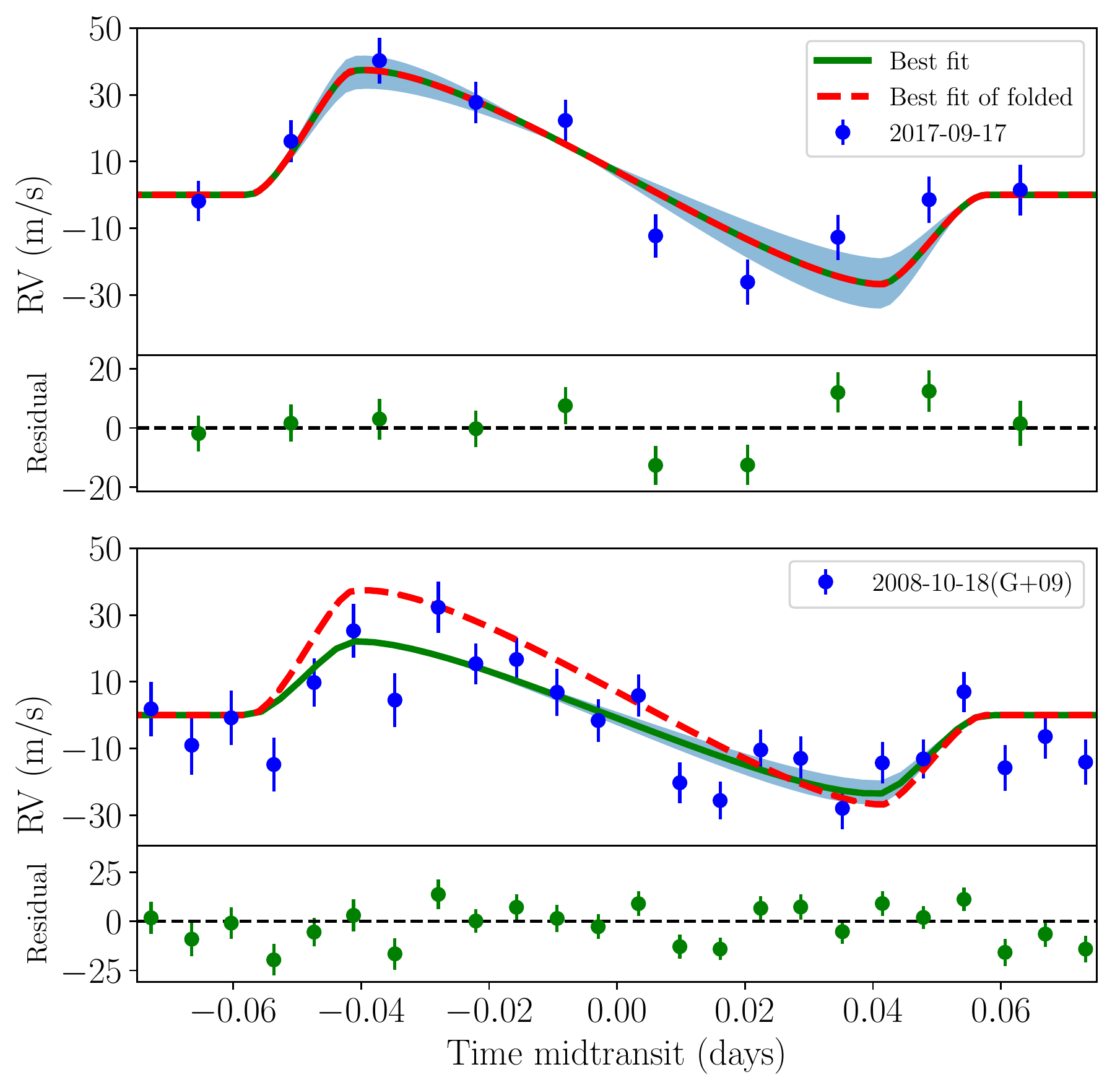}
\caption{As for Fig. C.1, but for WASP-6b. }
\end{figure}

\begin{figure}
\includegraphics[width=0.45 \textwidth, height=8cm]{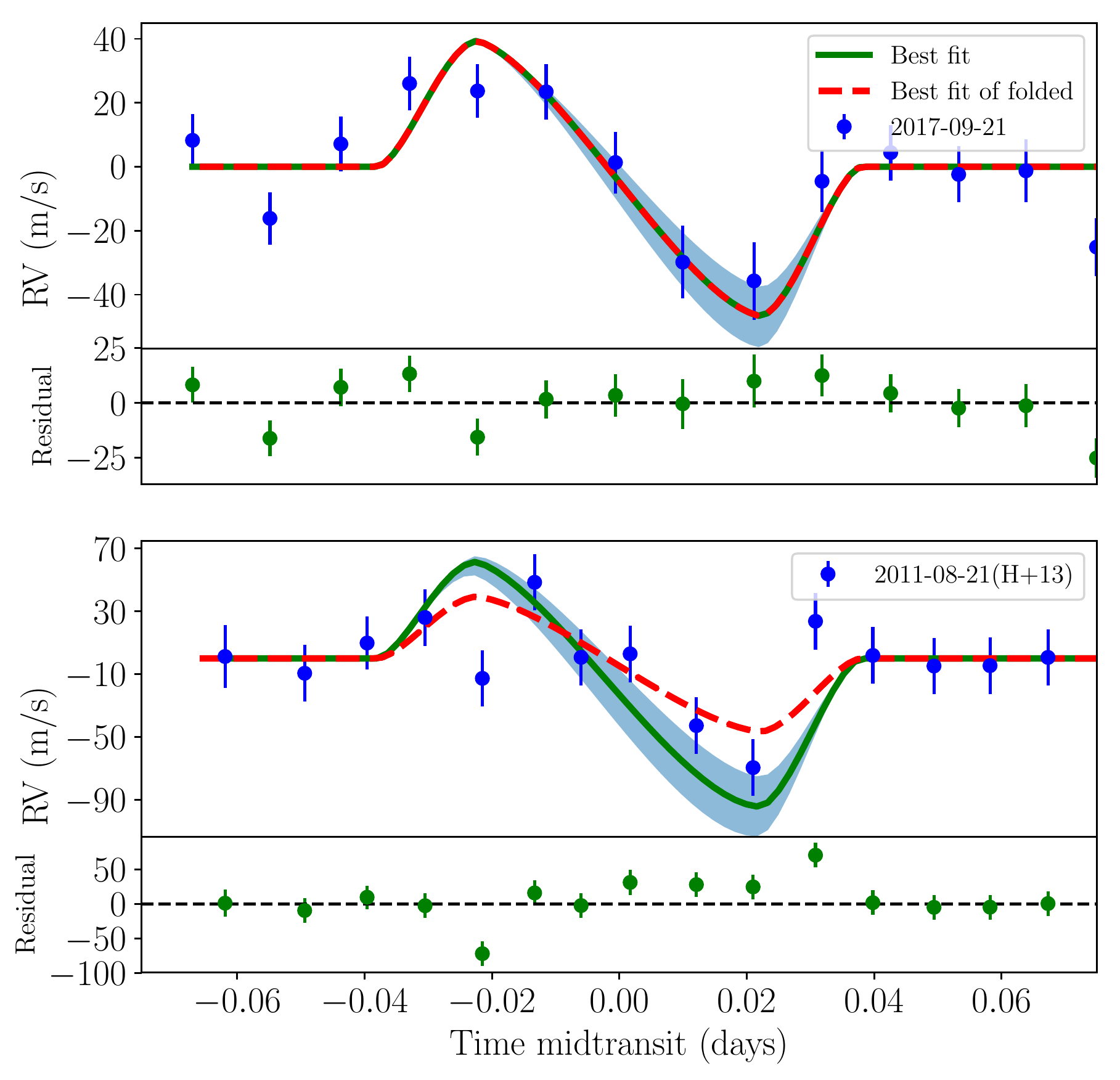}
\caption{As for Fig. C.1, but for WASP-52b. }
\end{figure}

\begin{figure}
\includegraphics[width=0.45 \textwidth, height=8cm]{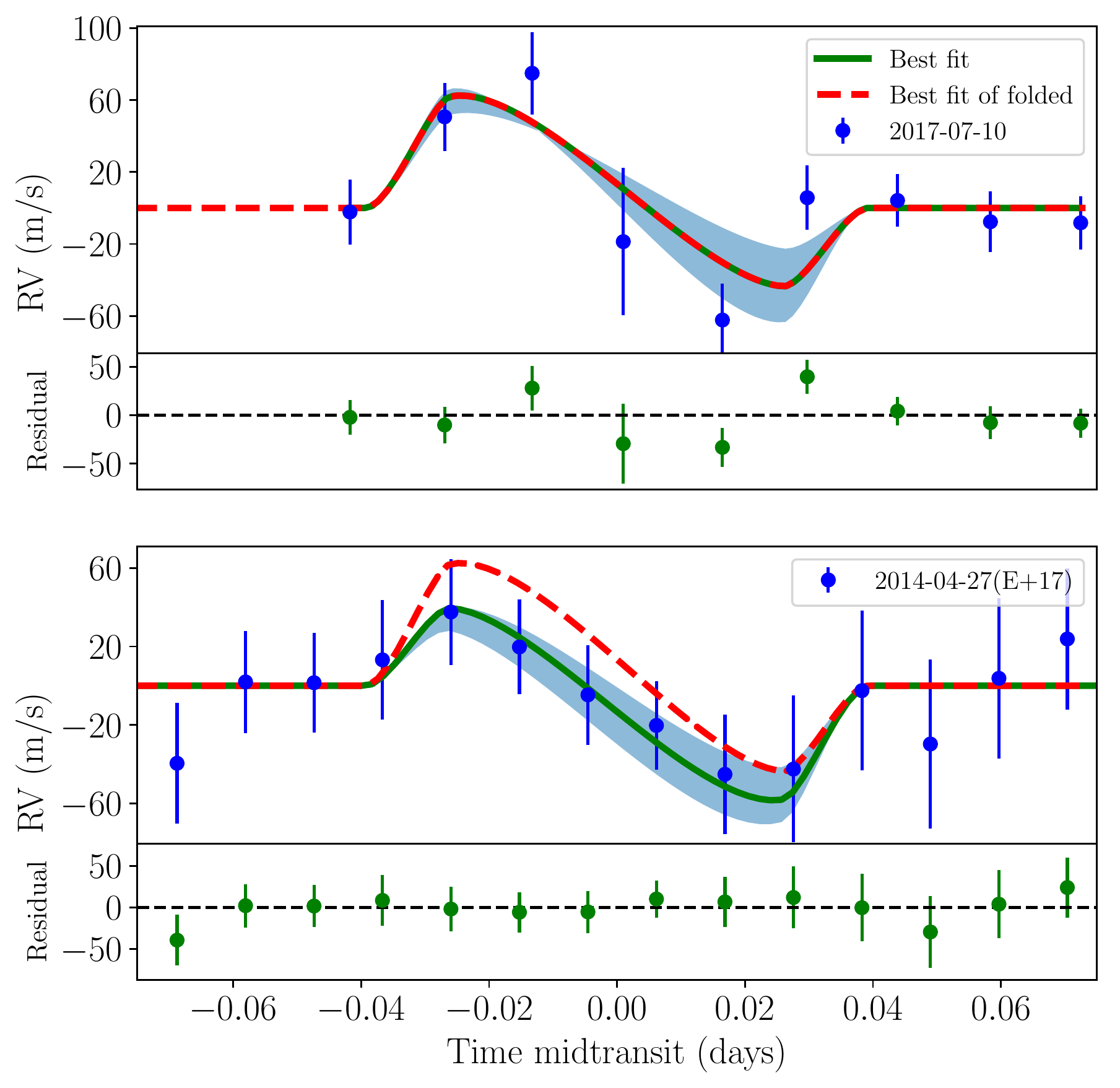}
\caption{As for Fig. C.1, but for Qatar-2b. }
\end{figure}

\section{RVs}
In this section we present our targets' RV measurements derived from the HARPS pipeline.
\begin{table}
\caption{RV measurements of CoRoT-2 derived from the HARPS pipeline.}              
\centering                                   
\begin{tabular}{c c c}          
\hline\hline                       
BJD-2400000 (days) & RV ($kms^{-1}$) & $\sigma$RV ($kms^{-1}$) \\
\hline 
57946.559605 & 23.23778 & 0.02662 \\
57946.570242 & 23.26527 & 0.02675 \\
57946.581503 & 23.20131 & 0.02707 \\
57946.592244 & 23.21717 & 0.02783 \\
57946.603506 & 23.47214 & 0.02658 \\
57946.613934 & 23.37828 & 0.0297 \\
57946.625300 & 23.24619 & 0.0387 \\
57946.636145 & 23.08137 & 0.02872 \\
57946.64699 & 22.88594 & 0.02873 \\
57946.658147 & 22.8558 & 0.03046 \\
57946.669097 & 22.88243 & 0.02795 \\
57946.679942 & 23.03788 & 0.026 \\
57946.690891 & 23.02822 & 0.02925 \\
57946.701840 & 23.03317 & 0.02887 \\
57946.712789 & 22.95865 & 0.03074 \\
57946.723634 & 22.90938 & 0.03362 \\
57953.524010 & 23.44181 & 0.04285 \\
57953.538894 & 23.31304 & 0.02706 \\
57953.555179 & 23.42134 & 0.02648 \\
57953.569288 & 23.46026 & 0.02304 \\
57953.583257 & 23.48048 & 0.02398 \\
57953.597922 & 23.38664 & 0.0236 \\
57953.611892 & 23.11847 & 0.02495 \\
57953.626278 & 23.01951 & 0.02395 \\
57953.640665 & 23.06027 & 0.02298 \\
57953.654634 & 23.16511 & 0.02677 \\
57953.669299 & 23.15558 & 0.02199 \\
57953.682991 & 23.11928 & 0.02284 \\
57953.697655 & 23.11298 & 0.02194 \\
58000.584972 & 23.43265 & 0.02276 \\
58000.599496 & 23.40732 & 0.02515 \\
58000.612771 & 23.34169 & 0.03296 \\
58000.627295 & 23.40437 & 0.0338 \\
58000.641542 & 23.59231 & 0.04754 \\
58000.655233 & 23.39131 & 0.04273 \\
58000.669897 & 23.14149 & 0.0482 \\
58000.68456 & 23.04224 & 0.06345 \\
58000.699223 & 22.94161 & 0.05071 \\
58000.713053 & 23.06745 & 0.04653 \\
58000.726745 & 23.05239 & 0.05873\\
\hline                                            
\end{tabular}
\end{table}

\begin{table}
\caption{Same as Table D.1, but for WASP-52.}              
\begin{tabular}{c c c}          
\hline\hline                       
BJD-2400000 (days) & RV ($kms^{-1}$) & $\sigma$RV ($kms^{-1}$) \\
\hline 
58017.583504 & -0.84985 & 0.00814 \\
58017.595657 & -0.87636 & 0.00816 \\
58017.606791 & -0.85499 & 0.00856 \\
58017.617578 & -0.83795 & 0.0084 \\
58017.628249 & -0.84215 & 0.00837 \\
58017.639025 & -0.84427 & 0.00872 \\
58017.649904 & -0.86827 & 0.00966 \\
58017.660471 & -0.90126 & 0.01142 \\
58017.671663 & -0.90908 & 0.01214 \\
58017.682334 & -0.87975 & 0.00957 \\
58017.693110 & -0.87265 & 0.00867 \\
58017.703781 & -0.88131 & 0.00876 \\
58017.714348 & -0.88202 & 0.0098 \\
58017.725436 & -0.9078 & 0.00907 \\
\hline                                            
\end{tabular}
\end{table}

\begin{table}
\caption{Same as Table D.1, but for WASP-41.}              
\begin{tabular}{c c c}          
\hline\hline                       
BJD-2400000 (days) & RV ($kms^{-1}$) & $\sigma$RV ($kms^{-1}$) \\
\hline 
57858.567723 & 3.31836 & 0.00337 \\
57858.579274 & 3.30996 & 0.00398 \\
57858.590153 & 3.30738 & 0.0039 \\
57858.600721 & 3.32713 & 0.0037 \\
57858.611392 & 3.32882 & 0.00376 \\
57858.622271 & 3.32647 & 0.00356 \\
57858.633047 & 3.31106 & 0.00359 \\
57858.643718 & 3.29118 & 0.00359 \\
57858.654494 & 3.26965 & 0.00376 \\
57858.665593 & 3.25685 & 0.00378 \\
57858.676160 & 3.26033 & 0.00341 \\
57858.686936 & 3.25868 & 0.0034 \\
57858.697723 & 3.28322 & 0.00333 \\
57858.708498 & 3.27431 & 0.00325 \\
57858.719273 & 3.26926 & 0.00305 \\
57858.730049 & 3.26569 & 0.003 \\
57864.679043 & 3.32545 & 0.00454 \\
57864.713684 & 3.35098 & 0.00407 \\
57864.724575 & 3.35114 & 0.00416 \\
57864.735246 & 3.31754 & 0.00403 \\
57864.746033 & 3.2992 & 0.00391 \\
57864.756797 & 3.28228 & 0.00376 \\
57864.767468 & 3.27314 & 0.00368 \\
57864.778452 & 3.26639 & 0.00386 \\
57864.789123 & 3.25747 & 0.00411 \\
57864.800003 & 3.284 & 0.00377 \\
57864.810570 & 3.28331 & 0.00416 \\
57864.821449 & 3.28484 & 0.00421 \\
57864.832329 & 3.28937 & 0.00424 \\
57864.843000 & 3.27784 & 0.00429 \\
\hline                                            
\end{tabular}
\end{table}

\begin{table}
\caption{Same as Table D.1, but for WASP-19.}              
\begin{tabular}{c c c}          
\hline\hline                       
BJD-2400000 (days) & RV ($kms^{-1}$) & $\sigma$RV ($kms^{-1}$) \\
\hline 
57878.566687 & 20.95255 & 0.00823 \\
57878.578411 & 20.95106 & 0.00772 \\
57878.589395 & 20.92428 & 0.00852 \\
57878.599961 & 20.92808 & 0.00805 \\
57878.611986 & 20.9094 & 0.01533 \\
57878.621928 & 20.89184 & 0.01078 \\
57878.63261 & 20.84059 & 0.01055 \\
57878.643073 & 20.77499 & 0.01215 \\
57878.653744 & 20.74375 & 0.01373 \\
57878.664519 & 20.77761 & 0.01815 \\
57878.675409 & 20.707 & 0.02424 \\
57878.686809 & 20.73201 & 0.01648 \\
57878.696959 & 20.69865 & 0.0147 \\
57878.707734 & 20.70051 & 0.01879 \\
57882.491266 & 20.97215 & 0.0113 \\
57882.502053 & 20.93926 & 0.01111 \\
57882.512723 & 20.9406 & 0.00884 \\
57882.52329 & 20.90364 & 0.01048 \\
57882.534169 & 20.88538 & 0.01184 \\
57882.544944 & 20.90923 & 0.01284 \\
57882.555835 & 20.89886 & 0.01419 \\
57882.566610 & 20.82783 & 0.014 \\
57882.577489 & 20.79082 & 0.01382 \\
57882.587951 & 20.75293 & 0.01453 \\
57882.598622 & 20.71149 & 0.01644 \\
57882.609501 & 20.76099 & 0.04641 \\
57882.620276 & 20.73152 & 0.01861 \\
57882.631260 & 20.67754 & 0.02311 \\
57882.641826 & 20.69615 & 0.02129 \\
57882.652509 & 20.66187 & 0.02625 \\
\hline                                            
\end{tabular}
\end{table}

\begin{table}
\caption{Same as Table D.1, but for WASP-6.}              
\begin{tabular}{c c c}          
\hline\hline                       
BJD-2400000 (days) & RV ($kms^{-1}$) & $\sigma$RV ($kms^{-1}$) \\
\hline 
58014.496106 & 11.48746 & 0.00749 \\
58014.509346 & 11.48237 & 0.00603 \\
58014.523767 & 11.49896 & 0.00629 \\
58014.537609 & 11.52177 & 0.0069 \\
58014.552690 & 11.50776 & 0.00621 \\
58014.566659 & 11.50096 & 0.00625 \\
58014.580768 & 11.46498 & 0.00653 \\
58014.595154 & 11.4497 & 0.00682 \\
58014.609262 & 11.46176 & 0.00679 \\
58014.623521 & 11.47164 & 0.00696 \\
58014.637769 & 11.47314 & 0.00763 \\
58014.652155 & 11.48487 & 0.00787\\
\hline                                            
\end{tabular}
\end{table}

\begin{table}
\caption{Same as Table D.1, but for Qatar-2.}              
\begin{tabular}{c c c}          
\hline\hline                       
BJD-2400000 (days) & RV ($kms^{-1}$) & $\sigma$RV ($kms^{-1}$) \\
\hline 
57945.459029 & -23.87141 & 0.01799 \\
57945.473819 & -23.8586 & 0.01898 \\
57945.487498 & -23.87141 & 0.02288 \\
57945.501744 & -24.00367 & 0.04094 \\
57945.517241 & -24.08928 & 0.02011 \\
57945.530503 & -24.05721 & 0.01787 \\
57945.544611 & -24.09699 & 0.01466 \\
57945.559135 & -24.1482 & 0.01674 \\
57945.573242 & -24.18702 & 0.01482\\
\hline                                            
\end{tabular}
\end{table}

\end{appendix}

\end{document}